\documentclass[12pt]{aastex63}
\usepackage{graphics,amsmath}
\usepackage{amsbsy}

\newcommand\eps{\epsilon}
\newcommand{\lapprox} {\, \lower3pt\hbox{$\sim$}\llap{\raise2pt\hbox{$<$}}\,}
\newcommand{\gapprox} {\, \lower3pt\hbox{$\sim$}\llap{\raise2pt\hbox{$>$}}\,}

\newcommand\mmatrix[1]{\textsf{\textbf{#1}}}
\renewcommand{\vec}[1]{ \protect {{\mathbf{\boldsymbol{#1}}}}}

\shorttitle{Anisotropic Radio-Wave Scattering}
\shortauthors{Kontar et al.}

\begin{document}

\title{Anisotropic Radio-Wave Scattering and the Interpretation of Solar Radio Emission Observations}

\author[0000-0002-8078-0902]{Eduard P. Kontar}
\affiliation{School of Physics \& Astronomy,
University of Glasgow,
Glasgow, G12 8QQ, UK}

\author[0000-0002-1810-6706]{Xingyao Chen} 
\affiliation{School of Physics \& Astronomy,
University of Glasgow,
Glasgow, G12 8QQ, UK}
\affiliation{Key Laboratory of Solar Activity, National Astronomical Observatories Chinese Academy of Sciences, Beijing 100101, China}

\author[0000-0002-4389-5540]{Nicolina Chrysaphi} 
\affiliation{School of Physics \& Astronomy,
University of Glasgow,
Glasgow, G12 8QQ, UK}

\author[0000-0001-6583-1989]{Natasha L.S. Jeffrey}
\affiliation{School of Physics \& Astronomy,
University of Glasgow, Glasgow, G12 8QQ, UK}
\affiliation{Northumbria  University,  Department  of  Mathematics,  Physics  and  Electrical  Engineering, Newcastle  upon  Tyne, NE1 8ST, UK}

\author[0000-0001-8720-0723]{A. Gordon Emslie}
\affiliation{Department of Physics \& Astronomy,
Western Kentucky University, KY 42101, USA}

\author[0000-0001-6185-3945]{Vratislav Krupar}
\affiliation{NASA Goddard Space Flight Center, Greenbelt, MD 20771, USA}
\affiliation{Institute of Atmospheric Physics CAS, Prague, Czech Republic}

\author[0000-0001-6172-5062]{Milan Maksimovic}
\affiliation{LESIA, Observatoire de Paris, Université PSL, CNRS, Sorbonne Université, Université de Paris, 5 place Jules Janssen, 92195 Meudon, France}

\author[0000-0003-2291-4922]{Mykola Gordovskyy}
\affiliation{School of Physics \& Astronomy, University of Manchester, Manchester M13 9PL, UK}

\author[0000-0002-7089-5562]{Philippa K. Browning}
\affiliation{Jodrell Bank Centre for Astrophysics, School of Physics \& Astronomy, University of Manchester, Manchester M13 9PL, UK }

\begin{abstract}

The observed properties (i.e., source size, source position, time duration, decay time) of solar radio emission produced through plasma processes near the local plasma frequency, and hence the interpretation of solar radio bursts, are strongly influenced by propagation effects in the inhomogeneous turbulent solar corona. In this work, a 3D stochastic description of the propagation process is presented, based on the Fokker-Planck and Langevin equations of radio-wave transport in a medium containing anisotropic electron density fluctuations. Using a numerical treatment based on this model, we investigate the characteristic source sizes and burst decay times for Type III solar radio bursts. Comparison of the simulations with the observations of solar radio bursts shows that predominantly perpendicular density fluctuations in the solar corona are required, with an anisotropy factor $\sim 0.3$ for sources observed at around 30~MHz. The simulations also demonstrate that the photons are isotropized near the region of primary emission, but the waves are then focused by large-scale refraction, leading to plasma radio emission directivity that is characterized by a half-width-half-maximum of about 40~degrees near 30~MHz. The results are applicable to various solar radio bursts produced via plasma emission.
\end{abstract}


\section{Introduction}

Solar radio emission is produced in the turbulent medium of the solar atmosphere, and its observed properties (source position, size, time profile, polarization, etc.) are significantly affected by the propagation of the radio waves from the emission site to the observer. Bright radio emission produced in the outer solar corona during flares is mostly produced via plasma emission mechanisms, so that the radiation is generated close to the plasma frequency or its harmonic \citep[see, e.g.,][for reviews]{1985srph.book..289S,2008A&ARv..16....1P}. Since the refractive index of an unmagnetized plasma $n_\text{ref}=(1-{\omega _{pe}^2}/{\omega ^2})^{1/2}$ is significantly different from unity for $\omega$ close to $\omega_{pe}$, the effects of density inhomogeneity along the wave path play a particularly strong role in the propagation of solar radio bursts produced by plasma processes. Appreciating this fact, even early observations \citep[e.g.,][]{1959IAUS....9..176W,1962AuJPh..15..180S,1971A&A....10..362S} considered radio-wave escape to be an important effect.

Scattering of radio waves on random density irregularities has long been recognized as an important process for the interpretation of radio source sizes \citep[e.g.,][]{1971A&A....10..362S},
positions \citep[e.g.,][]{1965BAN....18..111F,1972PASAu...2..100S}, directivity \citep[e.g.,][]{2007ApJ...671..894T,2008A&A...489..419B,2009SoPh..259..255R}, and intensity-time profiles \citep[e.g.,][]{2018ApJ...857...82K}. In the particularly strong scattering environment appropriate to electromagnetic waves close to the plasma frequency, the wave direction is quickly randomized, and the waves quickly become isotropic. As the waves propagate farther away from the source, large-scale refraction also produces a degree of focusing/defocusing. The observed properties of solar radio emission are therefore determined by an interconnected combination of scattering off small-scale inhomogeneities, which generally shifts the observed positions of sources {\it away} from the solar disk center \citep{1972PASAu...2..148R,2019ApJ...873...48G}, and refraction by relatively large-scale density inhomogeneities, such as Coronal Mass Ejection (CME) fronts \citep{2009AnGeo..27.3933A} or coronal streamers and fibers \citep{1977A&A....61..777B}), which generally shifts the sources {\it toward} the disk center \citep{1959IAUS....9..176W,1962AuJPh..15..180S,1971A&A....10..362S}.

Sub-arcminute imaging observations of Type III solar radio bursts have shown that intrinsic sources with sizes $\lapprox 0'.1$ result in observed sources as large as $\sim 20'$ at 30~MHz \citep{2017NatCo...8.1515K,2018SoPh..293..115S}, demonstrating that scattering dominates the properties of observed source sizes.  Moreover, the locations of the upper and lower sub-band sources of Type II solar radio bursts are observed to be spatially separated \cite[e.g.,][]{2012A&A...547A...6Z,2018ApJ...868...79C}, with the amount of separation being consistent with radio-wave scattering of plasma radio emission from a single region \citep{2018ApJ...868...79C}.

The majority of both past \cite[e.g.,][]{1971A&A....10..362S} and recent \cite[e.g.,][]{2008ApJ...676.1338T,2018ApJ...857...82K} ray-tracing simulations have assumed isotropic scattering by small-scale density fluctuations. However, there are observations \citep{1985srph.book..237M} that cannot be explained by the earlier models; for example, to provide a plausible explanation of the size and directivity of Type I solar radio bursts, a fibrous structure was invoked \citep{1977A&A....61..777B}. Other recent observations also suggest that the scattering is anisotropic, with the dominant effect being perpendicular to the heliospheric radial direction \citep{2017NatCo...8.1515K}.

A quantitative understanding of radio-wave propagation is particularly timely in the view of the opportunities to be opened by the Square Kilometer Array (SKA) \citep{2009IEEEP..97.1482D, 2019AdSpR..63.1404N} and the observations with the Chinese Spectral Radioheliograph \citep{2009EM&P..104...97Y,2016PASA...33...61L}. While there have been a number of  Monte Carlo simulations developed to describe wave scattering (mostly for isotropic density fluctuations), these do not all agree. Therefore, the present work addresses this important issue both by extending the isotropic plasma treatment of \cite{2019ApJ...873...33B} into the anisotropic scattering domain and by improving the previous descriptions by \cite{1971A&A....10..362S}, \cite{1999A&A...351.1165A}, and \cite{2008ApJ...676.1338T}.
The description presented captures both multiple scattering of radio waves in anisotropic small-scale turbulence and refraction of waves in the presence of large-scale plasma inhomogeneity.

In Section~\ref{sec:equations}, we present a general theoretical treatment of the scattering process, and apply it to both isotropic and (using a diagonalization scaling technique) axially-symmetric anisotropic scattering. In Section~\ref{sec:stochastic-equations}, we derive the pertinent stochastic differential equations that allow for a numerical solution for both isotropic and anisotropic turbulence. In Section~\ref{sec:results}, we review the numerical Monte Carlo technique used to solve Langevin equations modelling both source sizes and time profiles. In Section~\ref{sec:observations}, we review relevant observations of the variation of radio source sizes and decay times with frequency, and we compare these observations with our numerical solutions. This leads us to the conclusion that observations of Type III solar radio burst sizes and durations, over a broad range of frequencies, {\it require} anisotropic scattering, in the entire heliosphere between the Sun and the Earth, with an anisotropy factor of around 3--4 and with the density fluctuations predominantly perpendicular to the radial direction. As discussed in Section~\ref{sec:summary}, these observations provide essential density fluctuation anisotropy constraints for MHD turbulence models \citep{2010MNRAS.402..362S,2012ApJ...756...21Z}
over a wide range of locations between the Sun and the Earth.

\section{Radio-wave scattering equations}\label{sec:equations}

The propagation of radio waves in a turbulent medium can be effectively described using a kinetic approach \citep[e.g.,][]{1979A&A....73..292M,1999A&A...351.1165A,2019ApJ...873...33B}.
This approach describes the evolution of radio waves, in an inhomogeneous plasma with quasi-static density fluctuations, in the geometrical optics approximation \citep{1961wptm.book.....T,1978wpsr.book.....I}, i.e., when the scale length for variation of the wavelength $\lambda$ due to inhomogeneity is much smaller than the wavelength itself:

\begin{equation}\label{eq:geometric}
\left|\frac{\mathrm{d} \lambda}{\mathrm{d} r}\right| \ll 1 \,\,\, .
\end{equation}
This description ignores diffraction effects and is generally valid only for small amplitude density fluctuations \citep[e.g.][]{2012wop..book.....P}. Nevertheless, it adequately describes the multiple-scattering transport of radio waves with angular frequency $\omega$ (s$^{-1}$) near the local plasma frequency ${\omega}_{pe} (\vec{r}) = \sqrt{4\pi e^2 n (\vec{r})/m_e}$ (where $e$ and $m_e$ are respectively the electron charge [esu] and mass [g], and $n(\vec{r})$ [cm$^{-3}$] is the local plasma density) in the turbulent plasma of the solar atmosphere. Similar to the weak turbulence theory of Langmuir waves in a plasma \citep[][]{1995lnlp.book.....T}, such a description provides the basis for a statistical description of density and electromagnetic wave interactions. Since the group velocity of density fluctuations is much less than the speed of light, the density fluctuations can be treated as effectively static. Therefore, only elastic scattering conserving wavevector $|\vec{k}|$ of radio waves is considered. The description presented is also limited to an unmagnetized plasma environment \citep{1996ASSL..204.....Z}.

The spectral number density (or photon number) $N(\vec{k}, \vec{r},t)$ (cm$^{-3}$~[cm$^{-1}$]$^{-3}$) can be described in the geometric-optic approximation using a Fokker-Planck equation

\begin{equation}\label{eq:kin1}
\frac{\partial N}{\partial t} + \frac{d {\bf r}}{dt} \cdot \frac{\partial N}{\partial {\bf r}} + \frac{d\vec{k}}{dt} \cdot \frac{\partial N}{\partial{\vec k}} = \frac{\partial }{\partial{k_i}} \, D_{ij}\frac{\partial N}{\partial{k_j}} - \gamma \, N \,\,\, ,
\end{equation}
where $\int N(\vec k, \vec r) \, d^3 \vec k = N_0 (\vec r)$ [cm$^{-3}$] is the number density of photons, $k_i$ are Cartesian coordinates of wavevector $\vec{k}$ and the summation is understood for a repeated index, $i,j=1,2,3$.  ${d\vec{r}}/{dt}$, ${d\vec{k}}/{dt}$ are given by the Hamilton equations corresponding to the dispersion relation for electromagnetic waves in an unmagnetized plasma \citep{1963JATP...25..397H}:

\begin{equation}\label{eq:dr-vec_dt}
\frac{d\vec r}{dt}  = \vec {v}_g= \frac{\partial \omega}{\partial{\vec k}} = \frac{c^2}{\omega} \, \vec k \,\,\, ,
\end{equation}

\begin{equation}\label{eq:dk-vec_dt}
\frac{d\vec k}{dt}  = -\frac{\partial \omega}{\partial{\vec r}} = -\frac{\omega_{pe}}{\omega} \, \frac{\partial \omega_{pe}}{\partial{\vec r}} \,\,\, .
\end{equation}
Here the photon packet frequency in Equation~(\ref{eq:dr-vec_dt}) is found from the dispersion relation ${\omega}^2 = {\omega}_{pe}^2+c^2k^2$ for electromagnetic waves in an unmagnetized plasma,
and $\gamma$ [1/sec] is the collisional (free-free) absorption coefficient for radio waves in a plasma \citep[e.g.,][]{1981phki.book.....L}.

The diffusion tensor $D_{ij}$ appropriate to scattering \citep[cf.][]{1999A&A...351.1165A,2019ApJ...873...33B} is given by

\begin{equation}\label{eq:D}
D_{ij} = \frac { \pi \omega_{pe}^4} {4 \, \omega^2} \int q_i \, q_j \, S(\vec{q}) \, \delta(\vec q \cdot \vec {v}_g) \, \frac{d^3q}{(2\pi)^3} = \frac{\pi \omega_{pe}^4} {4 \, \omega c^2 } \int q_i \, q_j \,  S(\vec{q}) \, \delta(\vec q \cdot \vec {k}) \, \frac{d^3q}{(2\pi)^3} \,\,\, ,
\end{equation}
where $\vec{q}$ is the wavevector of electron density fluctuations. $S(\vec{q})$ is the spectrum of the density fluctuation normalized to the relative density fluctuation variance:

\begin{equation}\label{eq:spectr_norm}
\epsilon^2 = \frac{\langle\delta n^2\rangle}{n^2}= \int S(\vec{q}) \, \frac{d^3q}{(2\pi)^3} \,\,\, ,
\end{equation}
where $n=\langle n \rangle$ is the average plasma density, taken to be a slowly varying function of position. Note that Equations~(\ref{eq:D}) and~(\ref{eq:spectr_norm}) include a scaling of $(2\pi)^3$ in the definition of the spectral density $S({\bf q})$, consistent with the treatment of \cite{1999A&A...351.1165A}, but not with the scaling used by \cite{2019ApJ...873...33B}.

\subsection{Isotropic scattering}\label{sec:anisotropic-scattering}

The bulk of radio-wave scattering research has assumed an isotropic spectrum of density fluctuations: $S(\vec{q})=S(q)$. Such an assumption substantially simplifies the expression for the wave number diffusion tensor $D_{ij}$ (see Appendix \ref{app:iso} for details), so that Equation (\ref{eq:D}) becomes

\begin{equation}
\begin{split}\label{eq:D_iso}
D_{ij} & = \left(\delta _{ij}-\frac{k_i k_j}{k^2}\right) \frac{1}{32\pi} \, \frac {\omega_{p e}^{4}}{\omega c^{2}k} \int_{0}^{\infty}  q^3 \, S(q) \, dq  \\
& =\frac{\pi}{8} \, \frac{{\omega}^4_{pe}}{{\omega}c^2k} \, \bar{q} \, \frac{\langle \delta n^2\rangle}{n^2} \, \left(\delta_{ij}-\frac{k_i k_j}{k^2}\right)
= \frac{\nu_\text{s}k^2}{2} \, \left ( \delta_{ij}-\frac{k_i k_j}{k^2} \right ) \,\,\, ,
\end{split}
\end{equation}
where $\delta _{ij}$ is the Kronecker delta and we have introduced the spectrum-averaged mean wavenumber

\begin{equation}\label{eq:q_average}
\bar{q}=\frac{1} {\epsilon^2} \int q \, S(q) \, \frac{d^3q}{(2\pi)^3}
\end{equation}
and the scattering frequency

\begin{equation}\label{eq:nu_s}
\nu_{\rm s} = \frac{\pi}{4} \, \frac{{\omega}^4_{pe}}{{\omega}c^2k^3} \, \bar{q} \, \frac{\langle \delta n^2\rangle}{n^2} = \frac{\pi}{4} \, \frac{{\omega}^4_{pe}}{{\omega}c^2k^3} \, \bar{q} \, \epsilon^2 \,\,\, .
\end{equation}
Since the scattering frequency $\nu_{\rm s}$ is proportional to the spectrum-weighted mean wavenumber $\bar{q} \, \epsilon^2$, knowing this latter quantity leads to a determination of the scattering frequency. Equivalently, observations of radio-wave scattering in the solar corona provide a diagnostic of the level of density fluctuations via the quantity $\bar{q} \, \epsilon^2$. The assumption of isotropy of the scattering density fluctuations allows us to substantially simplify the diffusion operator, so that in spherical coordinates

\begin{equation}\label{d-spherical}
\frac{\partial }{\partial{k_i}} \left ( D_{ij}\frac{\partial }{\partial{k_j}} \right ) =
\frac{\partial }{\partial \mu} \left(\frac{\nu _s}{2} \left ( 1 - \mu^2 \right ) \, \frac{\partial }{\partial \mu}\right) \,\,\, ,
\end{equation}
where $\mu = \cos \theta$, $\theta$ being the polar angle for $\vec{k}$.

\subsection{Anisotropic scattering}

As we shall see below, using numerical simulations based on the isotropic scattering analysis above, isotropic scattering is inconsistent with the observations of solar radio source sizes and time profiles.  We therefore now develop a model for scattering in an anisotropic spectrum of density fluctuations $S(\vec q)$. Similar to previous investigations \cite[e.g.,][]{1968AJ.....73..972H}, we assume that the anisotropic density fluctuations are axially symmetric, so that the spectrum can be parameterized as a spheroid in $\vec{q}$-space:

\begin{equation}\label{eq:S_ani}
S(\vec q) = S \, \left( \left [ {q_{\perp}}^2+{\alpha}^{-2}{q_\parallel}^2 \right ]^{1/2}\right) \,\,\, ,
\end{equation}
where $\alpha =h_\perp/h_\parallel$ is the ratio of perpendicular and parallel correlation lengths (see also Appendix~\ref{app:gauss}). When $h_\perp \gg h_\parallel$ (i.e., $\alpha \gg 1$), the density fluctuations are mostly in the perpendicular direction; conversely, when $h_\perp \ll h_\parallel$ (i.e., $\alpha \ll 1$), the spectrum of density fluctuations is dominated by those parallel direction. For example, the direction parallel to heliospheric radial direction is following the guiding magnetic field in spherically symmetric corona.

It is convenient to introduce the anisotropy matrix

\begin{equation} \label{eq:A_matrix}
\mmatrix{A}=
{\left( \begin{array}{ccc}
1 & 0 & 0 \\
0 & 1 & 0 \\
0 & 0 & {\alpha}^{-1} \\
\end{array} \right )} \,\,\, .
\end{equation}
Then, defining $\vec{\widetilde q}= \mmatrix{A}\vec q$ (so that $\vec q= \mmatrix{A}^{-1} \vec{\widetilde q}$), we can write

\begin{equation}\label{q-qtilde}
{q_{\perp}}^2+{\alpha}^{-2}{q_\parallel}^2
= \vec{q} \, \mmatrix{A}^2 \, \vec{q} = q _i \, {A}^2_{ij} \, q_j=
 \vec{\widetilde q} \cdot \vec{\widetilde q} = {\widetilde q}_i{\widetilde q}_i\,\,\, ,
\end{equation}
where $q_{\perp}$ and $q_{\parallel}$ are respectively the perpendicular and parallel components of the wavevector $\vec{q}$.

Using Equations~(\ref{eq:D}), (\ref{eq:S_ani}), and~(\ref{q-qtilde}), the wave vector diffusion coefficient can be written as

\begin{equation}
\label{eq:Dij_aniso}
\begin{split}
D_{ij} &= \frac {\pi \omega_{pe}^4} {4\omega c^2 }
\int q_i \, q_j \, S(|\mmatrix{A} \vec q|) \, \delta(\vec q \cdot \vec k)  \, \frac{d^3q}{(2\pi)^3} \\
&= \frac {{ \omega } _ { p e } ^ { 4 } } { 32\pi^2 { \omega } c^2 } \int (\mmatrix{A}^{-1}\vec{\widetilde q})_i \, (\mmatrix{A}^{-1}\vec{\widetilde q})_j \, S(\vec{\widetilde q}) \, \delta({\vec{\widetilde q} \cdot \mmatrix{A}^{-1}\vec k}) \, d^3q \\
&= \frac {\omega_{pe}^4} {32\pi^2 { \omega } c^2 } \, {A}^{-1}_{i\alpha} \, {A}^{-1}_{i\beta} \int \vec{\widetilde q}_{\alpha} \,
\vec{\widetilde q}_{\beta} \, S(\vec{ q}) \, \delta(\vec{\widetilde q} \cdot \vec {\widetilde k}) \, \det(\mmatrix{J}) \, d^3\widetilde q \,\,\, ,
\end{split}
\end{equation}
where $\det(\mmatrix{J}) \equiv \det(\mmatrix{A}^{-1}) = \alpha$ is the determinant of the Jacobian matrix $\mmatrix{J}$ transforming coordinates from $\vec{q}$ to $\vec{\widetilde q}$.
Equation~(\ref{eq:Dij_aniso}) can be written as

\begin{equation}\label{eq:D_ijanis2}
D_{ij} =\frac {{ \omega } _ { p e } ^ { 4 } } { 32\pi^2 { \omega } c^2 } \, {A}^{-1}_{i\alpha} \, {A}^{-1}_{j\beta} \, \left ( {\delta}_{\alpha \beta}-\frac{{\widetilde k}_{\alpha} {\widetilde k}_{\beta}}{{\widetilde k}^2} \right ) \, \frac{\pi \alpha}{\widetilde k} \int _0^{\infty}{\widetilde q}^3 \, S(\widetilde q) \, d\widetilde q \,\,\, ,
\end{equation}
where we introduced $\vec {\widetilde k} = \mmatrix{A}^{-1} \vec k$.

We can now write the diffusion tensor components $D_{ij}$ in terms of the original quantities $\vec k$:

\begin{equation}
\label{eq:d_ij_aniso_k}
D_{ij} = \left[\frac{{A}^{-2}_{ij}}{(\vec k \mmatrix{A}^{-2} \vec k)^{1/2}}-\frac{(\mmatrix{A}^{-2}\vec k)_i(\mmatrix{A}^{-2}\vec k)_j}{(\vec k \mmatrix{A}^{-2} \vec k)^{3/2}}\right]
\frac {{ \omega } _ { p e } ^ { 4 } } { 32\pi { \omega } c^2 } \alpha \int_0^{\infty} {\widetilde q}^3 \, S(\widetilde q) \, d\widetilde q \,\,\, .
\end{equation}
For isotropic scattering, the anisotropy matrix $\mmatrix{A}$ reduces to the identity matrix and Equation~(\ref{eq:d_ij_aniso_k}) correspondingly reduces to Equation~(\ref{eq:D_iso}).
Equation~(\ref{eq:d_ij_aniso_k}) coincides with Equation (B10) of \cite{1999A&A...351.1165A} (note the equation sign misprint in their paper appendix).

\section{Stochastic differential equations}\label{sec:stochastic-equations}

We now proceed to cast the Fokker-Planck equation~(\ref{eq:kin1}) in a form suitable for numerical computation.  The scattering term in Equation~(\ref{eq:kin1}) can be written as

\begin{equation}\label{eq: D_ito}
\frac{dN}{dt}=\frac{\partial}{\partial k_i} D_{ij} \frac{\partial N}{\partial k_j} =
\frac{\partial}{\partial k_i} \left( - N \, \frac{\partial  D_{ij}}{\partial k_j} + \frac{\partial}{\partial k_j} D_{ij} \, N \right ) =
\frac{\partial}{\partial k_i} \left( - N \, \frac{\partial  D_{ij}}{\partial k_j} +\frac{\partial}{\partial k_j} \frac{1}{2}B_{im}B_{jm}^{\text{T}} \, N \right ) \,\,\, ,
\end{equation}
where $\mmatrix{B}$ is a positive-semi-definite matrix with matrix elements determined by matrix $D$, so that

\begin{equation}\label{eq: D_BB}
D_{ij}= \frac{1}{2}B_{im}B_{jm}^{\text{T}} \,\,\, .
\end{equation}
The nonlinear Langevin equation for $\vec{k}(t)$ corresponding to the Fokker-Planck equation (\ref{eq: D_ito}) is

\begin{equation}\label{eq:ito1}
\frac{dk_i}{dt}=\frac{\partial  D_{ij}}{\partial k_j}+ B_{ij} \, {\xi}_j \,\,\, ,
\end{equation}
where $\vec{\xi} (t)$ is a Gaussian white noise with  the properties $\langle \vec{\xi} (t) \rangle = 0$ and $\langle \xi _i(0) \, \xi _j(t) \rangle=\delta_{ij} \, \delta (t)$, where $\langle ...\rangle$ denotes an ensemble average, $\delta (t)$ is the Dirac delta function, and the $\vec{k}$-dependent  deterministic vectors $\partial  D_{ij}/\partial k_j$ and $B_{ij}$ correspond to the diffusion tensor $D_{ij}$. These are analogous to the equations describing binary collisions in a plasma \citep[see, e.g.,][]{1978DoSSR.238.1324I,1979TMP....39..456S,2014JCoPh.274..140R}.
Equation~(\ref{eq:ito1}) is similar to the equation by \cite{1999A&A...351.1165A}; it is the definition of the stochastic integral in It\^{o}'s sense, adopted in the theory of random processes. It\^{o}' approach considerably simplifies its numerical integration and requires the knowledge of  function $D_{ij}(\vec{k})$ at the beginning of the time step rather than half-step in Stratonovich form \citep{1978DoSSR.238.1324I}. The first term on the RHS describes the so-called It\^{o} drift, a systematic decrease of $k_i$ due to elastic scattering, while the second term represents diffusion. The presence of the It\^{o} drift improves the stochastic differential equations used in the past \cite[e.g.,][]{1971A&A....10..362S,1974SoPh...35..153R,2007ApJ...671..894T} and conserves the value of $|\vec{k}|$ for elastic scattering.

If we apply It\^{o}'s formula to the square of the wavevector $\vec{k}\cdot\vec{k}=k_ik_i$, one finds

\begin{equation}\label{eq:k2_const}
\frac{d}{dt} \, k_ik_i = 2k_i\frac{dk_i}{dt}+B_{ij}B_{ij}=0 \,\,\, ,
\end{equation}
where we have used $k_i \, dk_i/dt = - k_i \, {\partial  D_{ij}}/{\partial k_j} = - \nu_\text{s}k^2$ and $B_{ij}B_{ij}=2 \nu_\text{s}k^2$. One can see that the presence of the so-called It\^{o} drift is necessary to ensure conservation of $k=|\vec{k}|$ in scattering events, similar to pitch angle scattering in a Lorentz gas \cite[e.g.,][]{1978DoSSR.238.1324I}.

Including large-scale refraction due to gradual variation of the ambient density $n(\vec{r})$ of the solar corona, the equation for the components of wavevector $\vec{k}$ becomes

\begin{equation}\label{eq:dk_dt_all}
\frac{dk_i}{dt} = -\frac{\omega_{pe}}{\omega} \, \frac{\partial \omega_{pe}}{\partial{r}} \, \frac{r_i}{r} + \frac{\partial D_{ij}}{\partial k_j}+ B_{ij} \, {\xi}_j \,\,\, ,
\end{equation}
which, in combination with the radio-wave transport equation

\begin{equation}\label{eq:dr_dt}
\frac{dr_i}{dt}  =\frac{c^2}{\omega} \, k_i \,\,\, ,
\end{equation}
describes the propagation, refraction, and scattering of radio-wave packets in an inhomogeneous plasma.

\subsection{Numerical solution of the Langevin equations}\label{sec:numerical-solution}

Following the conceptually similar description of plasma collisions, we modify the transport code of \cite{2014ApJ...787...86J}, giving the wave-vector and position of photons at the next time step from the stepping equations

\begin{equation}\label{eq:photon}
\begin{split}
k_i(t+\Delta t) & = k_i (t) - \frac{\omega_{pe} (r(t))}{\omega} \, \frac{\partial \omega_{pe}}{\partial{r}} (t) \, \frac{r_i(t)}{r(t)} \, \Delta t +
\frac{\partial  D_{ij}}{\partial k_j}\, \Delta t + B_{ij} \, {\xi}_j \, \sqrt{\Delta t} \,\,\, , \\
r_i(t+\Delta t) &= r_i(t) + \frac{c^2}{\omega} \, k_i(t) \, \Delta t \,\,\, ,
\end{split}
\end{equation}
where the ${\xi}_i$ are random numbers drawn from the normal distribution $N(0,1)$ with zero mean and unit variance.

\subsubsection{Isotropic scattering}

In the case of isotropic density fluctuations (and hence isotropic scattering), the Langevin equations take on a particularly simple form. With $D_{ij}$ now given by Equation~(\ref{eq:D_iso}), one finds that

\begin{equation}\label{eq:D_ij_dkIso}
  \frac{\partial  D_{ij}}{\partial k_j} = -2 \, \frac{\pi}{8} \, \frac{{\omega}^4_{pe}}{{\omega}c^2k} \, \bar{q} \, \frac{\langle \delta n^2\rangle}{n^2} = - \nu_{\rm s} \, k_i
\end{equation}
and

\begin{equation}\label{eq:B_ij_iso}
B_{ij} = \left(\frac{\pi}{4}
\frac{{\omega}^4_{pe}}{{\omega}c^2k} \, \bar{q} \, \frac{\langle \delta n^2\rangle}{n^2} \right)^{1/2}\left(\delta_{ij}-\frac{k_i k_j}{k^2}\right)
=\sqrt{{\nu _sk^2}} \left(\delta_{ij}-\frac{k_i k_j}{k^2}\right) \,\,\, ,
\end{equation}
so that Equations~(\ref{eq:photon}) become

\begin{equation}
\label{eq:k_i_t_dt}
k_i(t+\Delta t) =k_i(t)-\frac{\omega_{pe}(r(t))}{\omega}\frac{\partial \omega_{pe}}{\partial r} (t) \, \frac{r_i(t)}{r(t)} \, \Delta t - \nu_{\rm s} \, k^2 \, \frac{k_i}{k^2} \, \Delta t + (\nu_\text{s} k^2 )^{1/2} \left ( \xi_i-\frac{k_i
(\vec{k} \cdot \vec{\xi})}{k^2} \right ) \, (\Delta t)^{1/2}
\end{equation}
and

\begin{equation}
\label{eq:r_i_t_dt}
r_i(t+\Delta t) =r_i(t) + \frac{c^2}{\omega} \, k_i(t) \,  \Delta t \,\,\, ,
\end{equation}
where, again, $\vec{\xi}$ is a vector with components $\xi _i$ being random numbers drawn from the normal distribution $N(0,1)$. Equations~(\ref{eq:k_i_t_dt}) and~(\ref{eq:r_i_t_dt}) are the Euler-Maruyama approximation to the Langevin equations~(\ref{eq:dk_dt_all}) and~(\ref{eq:dr_dt}); they are in a form particularly useful for solving initial value problems. The time step $\Delta t$ is chosen to be much smaller than the characteristic times due to scattering and refraction. The mean scattering time $1/\nu_\text{s}$ is normally the smaller time, and so we choose $\Delta t = 0.1/\nu _\text{s}$. Since $\nu_\text{s} (r)$ is a decreasing function of $r$, the time step is shortest near the radio emission source and quickly increases with distance.

\subsection{Anisotropic scattering}

Now let us find the Langevin equation functions for the anisotropic scattering tensor given by Equation~(\ref{eq:d_ij_aniso_k}). For the anisotropy matrix $\mmatrix{A}$ given by Equation (\ref{eq:A_matrix}),

\begin{equation}\label{eq:d_kj_kAk}
\frac{\partial}{\partial k_j} (\vec k \mmatrix{A}^{-2} \vec k)=2(\mmatrix{A}^{-2}\vec{k})_j = 2{A}^{-2}_{ji}k_i \,\,\, ,
\end{equation}

\begin{equation}\label{eq:dkj_kA2k1.5}
 \frac{\partial}{\partial k_j} (\vec k \mmatrix{A}^{-2} \vec k)^{3/2}
=3 (\vec k \mmatrix{A}^{-2} \vec k)^{1/2}(\mmatrix{A}^{-2}\vec{k})_j
=3 \widetilde k {A}^{-2}_{ji}k_i
\,\,\, ,
\end{equation}
and

\begin{equation}\label{eq:d_kj_kAk05}
\frac{\partial}{\partial k_j} \frac{1}{(\vec k \mmatrix{A}^{-2} \vec k)^{1/2}}
= -\frac{(\mmatrix{A}^{-2}\vec k)_j}{(\vec k \mmatrix{A}^{-2} \vec k)^{3/2}}
= - \frac{{A}^{-2}_{ji}k_i}{\widetilde k ^3} \,\,\, ,
\end{equation}
where the ${A}^{-2}_{ji}$ are elements of the diagonal matrix $\mmatrix{A}^{-2}=\mmatrix{A}^{-1}\mmatrix{A}^{-1}$ and summation over repeated indices is implicit. Using the definitions $\vec {\widetilde k}=\mmatrix{A}^{-1} \vec k$ and $\widetilde k=|\vec {\widetilde k}|=|\vec k \mmatrix{A}^{-2} \vec k|^{1/2}$, one finds the explicit expressions for Langevin equations in case of anisotropic scattering:

\begin{equation}
\begin{split}
\label{eq:dD_dk_ani}
\frac{\partial}{\partial k_j} D_{ij} &=D_A
\left[-\frac{{A}^{-2}_{ij}(\mmatrix{A}^{-2}\vec{k})_j}
{{(\vec k \mmatrix{A}^{-2}\vec{k})}^{3/2}}
-\frac{{A}^{-2}_{ij}(\mmatrix{A}^{-2}\vec{k})_j+{A}^{-2}_{jj}(\mmatrix{A}^{-2}\vec{k})_i }{{(\vec k \mmatrix{A}^{-2}\vec{k})}^{3/2}}
+ 3 \, \frac{(\mmatrix{A}^{-2}\vec{k})_i(\mmatrix{A}^{-2}\vec{k})_j(\mmatrix{A}^{-2}\vec{k})_j}{{(\vec k \mmatrix{A}^{-2}\vec k)}^{5/2}}\right] \\
&=D_A
\left[-\frac{(\mmatrix{A}^{-4}\vec{k})_i}{{\widetilde k}^3}
-\frac{(\mmatrix{A}^{-4}\vec{k})_i+\mbox{tr}(\mmatrix{A}^{-2})(\mmatrix{A}^{-2}\vec{k})_i}{{\widetilde k}^3}+\frac{3 (\mmatrix{A}^{-2}\vec{k})_i (\vec k \mmatrix{A}^{-4}
\vec k)}{{\widetilde k}^5}\right] \\
&= \frac{D_A}{{\widetilde k}^5}\left[-2{\widetilde k}^2(\mmatrix{A}^{-4}{\vec k})_i+(\mmatrix{A}^{-2}\vec {k})_i
\left( 3(\vec k \mmatrix{A}^{-4} \vec{k})- (2+{\alpha}^2){{\widetilde k}}^2\right)\right ] \,\,\, ,
\end{split}
\end{equation}
where $\mbox{tr}(\mmatrix{A}^{-2})=2+\alpha^2$ is the trace of matrix $\mmatrix{A}^{-2}$ for the anisotropy matrix $\mmatrix{A}$ given by Equation~(\ref{eq:A_matrix}), and

\begin{equation}\label{eq:B_ijanis}
B_{ij}=\sqrt{\frac{2D_A}{(\vec k \mmatrix{A}^{-2} \vec k)^{1/2}}} \, {A}^{-1}_{i\alpha} \left[{\delta}_{\alpha j}-\frac{({\mmatrix{A}^{-1}\vec{k}})_{\alpha} ({\mmatrix{A}^{-1}\vec{k}})_{j}}{{\widetilde k}^2}\right]
=\sqrt{\frac{2D_A}{\widetilde k}} \left[A^{-1}_{ij}-\frac{({\mmatrix{A}^{-2}\vec{k}})_i ({\mmatrix{A}^{-1}\vec{k}})_j}{{\widetilde k}^2}\right] \,\,\, ,
\end{equation}
where

\begin{equation}\label{eq:D_A}
  D_A= \frac {{ \omega } _ { p e } ^ { 4 } } { 32\pi { \omega } c^2 } \alpha \int_0^{\infty} {\widetilde q}^3 \, S(\widetilde q) \, d\widetilde q
\end{equation}
is the $k$-independent coefficient in the diffusion tensor~(\ref{eq:d_ij_aniso_k}). The Langevin equations (\ref{eq:dk_dt_all}), together with the vector functions (\ref{eq:dD_dk_ani}) and~(\ref{eq:B_ijanis}), can be solved numerically for an arbitrary spectrum of density fluctuations.  For isotropic scattering, i.e., in the limit $\alpha =1$, the functions (\ref{eq:dD_dk_ani}) and~(\ref{eq:B_ijanis}) reduce to Equations~(\ref{eq:D_ij_dkIso}) and~(\ref{eq:B_ij_iso}), respectively.

Due to the choice of anisotropy matrix (Equation~(\ref{eq:A_matrix})), it is useful to introduce the perpendicular

\begin{equation}
\begin{split}
\label{eq:A_perp}
\frac{\partial}{\partial k_j} D_{\perp \,j}&=
\frac{D_A k_{\perp}}{{\widetilde k}^5}
\left[-2{\widetilde k}^2+(1-{\alpha}^2){\widetilde k}^2
+3{\alpha}^2({\alpha}^2-1){k_{\parallel}}^2\right] \\
&=\frac{D_A k_{\perp}}{{\widetilde k}^5}\left[-(1+{\alpha}^2){\widetilde k}^2
+3{\alpha}^2({\alpha}^2-1){k_{\parallel}}^2\right]
\end{split}
\end{equation}
and parallel

\begin{equation}
\begin{split}
\label{eq:A_par}
\frac{\partial}{\partial k_j} D_{\parallel \,j}
&=\frac{D_A k_{\parallel}}{{\widetilde k}^5}
\left[-2{\widetilde k}^2{\alpha}^4+{\alpha}^2\left[(1-{\alpha}^2){\widetilde k}^2
+3{\alpha}^2({\alpha}^2-1)k_{\parallel}^2\right]\right] \\
&=\frac{D_A k_{\parallel}}{{\widetilde k}^5}
\left[(-3{\alpha}^4+{\alpha}^2)
{\widetilde k}^2+3{\alpha}^4({\alpha}^2-1)k_{\parallel}^2\right]
\end{split}
\end{equation}
components of ${\partial} D_{i j}/{\partial k_j}$ in the differential Equation~(\ref{eq:dk_dt_all}). These equations differ from Equation~(44) in \citet{1999A&A...351.1165A}. For isotropic scattering ($\alpha =1$), Equations~(\ref{eq:A_perp}) and~(\ref{eq:A_par}) reduce to the isotropic case. The expressions  (\ref{eq:A_perp}) and~(\ref{eq:A_par}) remain finite for the limiting cases of quasi-perpendicular density fluctuations, i.e., $\alpha \rightarrow \infty$, as well as in the quasi-longitudinal case $\alpha \rightarrow 0$.

One can also readily verify the result~(\ref{eq:B_ijanis}) {\it a posteriori}:

\begin{equation}
\begin{split}
\label{eq-anis14}
\frac{1}{2} \, B_{ik}\times B^T_{jk}&=D_A \, {A}^{-1}_{ik} \, {A}^{-1}_{kj} \left[{\delta}_{i k}-\frac{({\mmatrix{A}^{-1}\vec{k}})_i ({\mmatrix{A}^{-1}\vec{k}})_k}{{\widetilde k}^2}\right]\times \left[{\delta}_{kj}-\frac{({\mmatrix{A}^{-1}\vec{k}})_k ({\mmatrix{A}^{-1}\vec{k}})_j}{{\widetilde k}^2}\right] \\
&=D_A \, {A}^{-1}_{ik} \, {A}^{-1}_{kj}\left[{\delta}_{i j}-\frac{({\mmatrix{A}^{-1}\vec{k}})_i ({\mmatrix{A}^{-1}\vec{k}})_j}{{\widetilde k}^2}
-\frac{({\mmatrix{A}^{-1}\vec{k}})_i ({\mmatrix{A}^{-1}\vec{k}})_j}{{\widetilde k}^2}
+\frac{{\widetilde k}^2 ({\mmatrix{A}^{-1}\vec{k}})_i  ({\mmatrix{A}^{-1}\vec{k}})_j}{{\widetilde k}^4}\right]\\
&=D_A\left[{A}^{-2}_{ij}-\frac{({\mmatrix{A}^{-2}\vec{k}})_i ({\mmatrix{A}^{-2}\vec{k}})_j}{{\widetilde k}^2}\right]\\
&=D_{ij} \,\,\, ,
\end{split}
\end{equation}
as required. We note that the ``square root'' of a matrix is not unique, and so, to simplify the numerical solution of the equations, we follow \citet{2011JChPh.135h4116S} in the choice for $B_{ij}$.

\subsection{Collisional absorption of radio waves}\label{sec:absorption}

The plasma of the solar corona is a collisional medium, which leads to free-free absorption of propagating electromagnetic waves, with a characteristic rate $\gamma$.  For binary collisions in a plasma \citep[e.g.,][]{1981phki.book.....L,1980panp.book.....M},

\begin{equation}\label{eq:gamma}
  \gamma = \frac{\omega_{pe}^2}{\omega^2} \, \gamma_{\mathrm{c}} \,\,\, ,
\end{equation}
where

\begin{equation}\label{eq:gamma_c}
  \gamma_{\mathrm{c}} = \frac{4}{3}\sqrt{\frac{2}{ \pi}} \, \frac{e^4 n(\vec{r}) \ln \Lambda }{m \, v_{\rm{Te}}^{3}} \,\,\, .
\end{equation}
Here the thermal speed $v_{\rm {Te}} = \sqrt{T_{e} / m_{e}}$, with $T_e$ the electron temperature in energy units.  A constant Coulomb logarithm $\ln \Lambda\simeq 20$ is assumed, per \cite{2014A&A...562A..57R}. We also assume an isothermal solar corona with temperature $T=86$~eV.

The effects of collisional absorption are stronger in higher density plasmas. The attenuation of the signal due to absorption is given by

\begin{equation}\label{eq:N_t}
N(t) = N_0 \, e^{- \tau_a} \,\,\, ,
\end{equation}
where the Coulomb collisional depth

\begin{equation}\label{eq:tau_a}
\tau_a = \int \gamma (\vec{r}(t)) \, dt \,\,\, .
\end{equation}
Absorption is in general always important at higher frequencies $\gtrsim 50$~MHz and noticeably affects the time profiles at higher frequencies. The effect of absorption is also noticeable when the scattering is so strong that the photons are trapped near the source for the time longer than free-free absorption time $1/\gamma$.

\section{Monte Carlo Ray-Tracing Simulations}\label{sec:results}

\subsection{Methodology}\label{sec:methodology}

\begin{figure}[pht]
\centering
\includegraphics[width=0.6\linewidth,keepaspectratio=true]{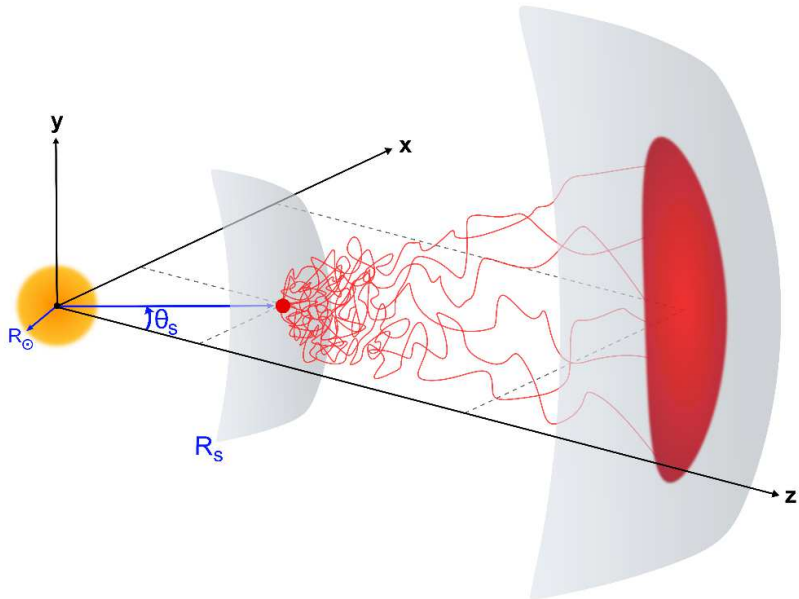}
\caption{\label{fig:cartoon} Cartoon showing the Sun-centered cartesian coordinate system $(x,y,z)$, where the $z$-axis is directed towards the observer. The initial location of a point source of radio emission is given by the radial coordinate $R_{s}$ and the polar angle $\theta _{s}$; the azimuth angle in the plane of the sky is not relevant to our study.  The photons scatter until they cross a sphere at a distance large enough that scattering is no longer important, resulting in an apparent source size and position indicated by the red region.}
\end{figure}

We have simulated the propagation of radio waves in the presence of background density fluctuations, using the Monte Carlo ray-tracing method presented in Section \ref{sec:stochastic-equations} (Equations~(\ref{eq:photon})). Simulations were performed in the solar centered coordinate system $(x,y,z)$ as shown in Figure \ref{fig:cartoon}, with the $z$-axis directed towards the observer; $x$ and $y$ are heliocentric-cartesian coordinates in the plane of the sky, used in solar imaging observations \citep{2006A&A...449..791T}.

\begin{figure}[pht]
\centering
\includegraphics[width=0.6\linewidth,keepaspectratio=true]{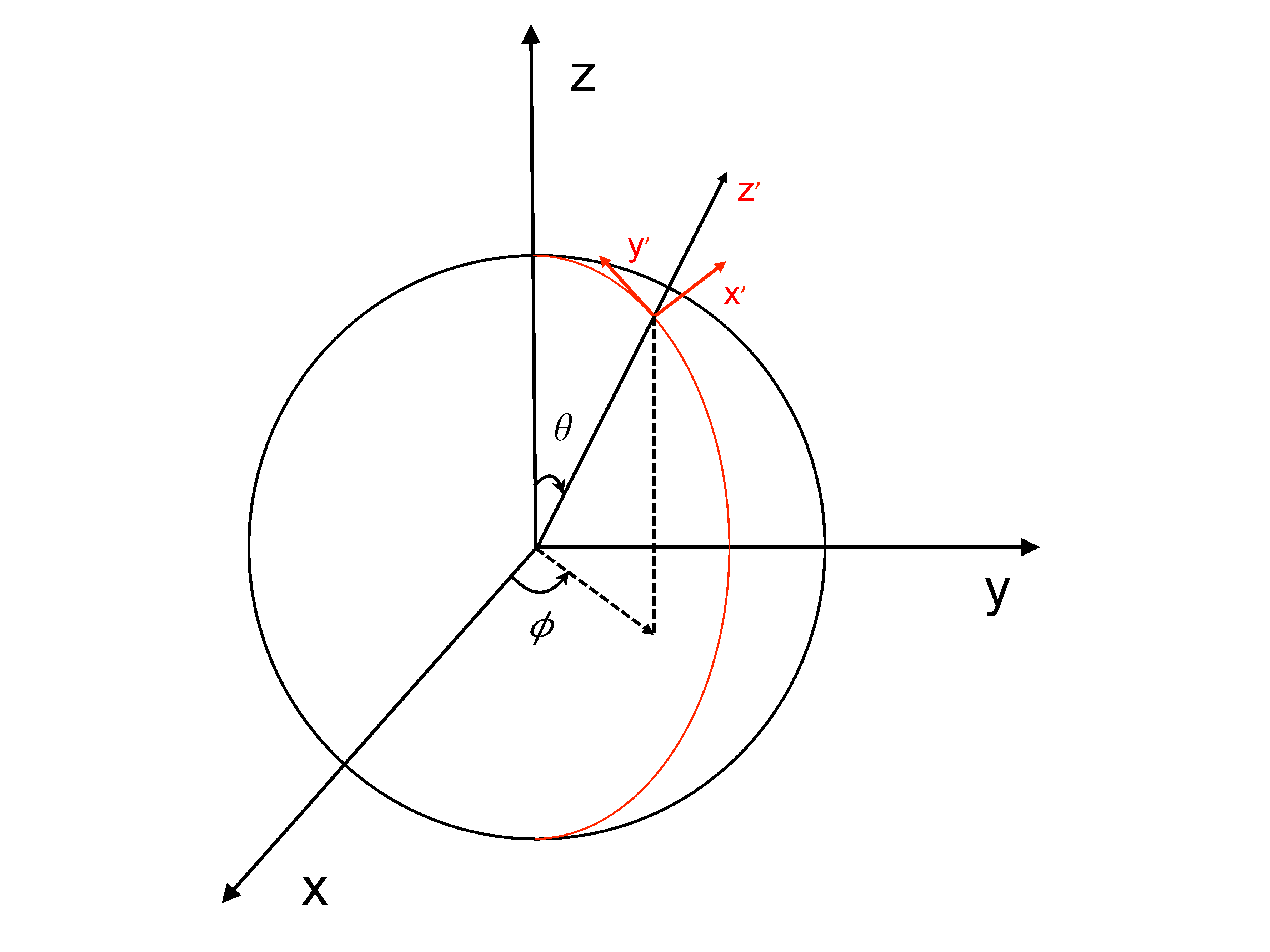}
\caption{\label{fig:q2xyz} Coordinate systems $(x,y,z)$ and $(x',y',z')$ with the Sun centre in the origin, where $z$-axis is directed to an observer and the $z'$-axis is parallel to $\vec r$, and the $y'$-axis is tangent to the circle created by the intersection of the plane formed by the $z$ and $z'$ axes and a spherical surface of radius $r$.}
\end{figure}

The solar corona is assumed to be spherically symmetric and the density fluctuations are assumed to be aligned with respect to the local radial direction, so that $q_{\parallel}$ is parallel to $\vec{r}$ for a given photon location. Similar to \citet{2010A&A...513L...2K} and \citet{2011A&A...536A..93J}, before advancing the stochastic differential equations~(\ref{eq:dD_dk_ani}) and~(\ref{eq:B_ijanis}) corresponding to the Langevin equations~(\ref{eq:dk_dt_all}), the wavevector $\vec{k}$ is first rotated to a local $(x',y',z')$ coordinate system where $z'$ is radially aligned (see Figure~\ref{fig:q2xyz}). In the paper, we only consider spherically symmetric solar corona\footnote{The approach can include arbitrary alignment and hence trace the local density anisotropy given by, e.g., a magnetic field.}. The stochastic differential equations are then advanced one time step and then the wavevector $\vec{k}$ is rotated back to the fixed $(x,y,z)$ coordinate system for propagation to the next scattering event. Figure~\ref{fig:q2xyz} shows the corresponding geometry; the $z'$-axis is parallel to $\vec r$, and the $y'$-axis is tangent to the circle created by the intersection of the plane formed by the $z$ and $z'$ axes and a spherical surface of radius $r$. The relationships between the wavevector components are

\begin{equation}
\begin{split}
\label{eq-0501}
k_x & = - k_{\perp x } \sin\phi+(k_{\parallel } \sin\theta-k_{\perp y } \cos\theta) \cos\phi \\
k_y & = k_{\perp x } \cos\phi+(k_{\parallel } \sin\theta-k_{\perp y } \cos\theta) \sin\phi \\
k_z & = k_{\parallel } \cos\theta+k_{\perp y } \sin\theta \,\,\, ,
\end{split}
\end{equation}
where $(k_x, k_y, k_z)$ are the components in the $(x,y,z)$ coordinate system, $(k_{\perp x }, k_{\perp y}, k_{\parallel})$ are the components in the $(x',y',z')$ coordinate system, and the rotation angles are given by the photon position in the $(x,y,z)$ coordinate system,

\begin{equation}\label{eq: angles}
\tan \phi =y/x\,,\;\;\; \sin\theta =\sqrt{1-z^2/r^2}\,,\;\;\;\cos\theta =z/r \,\,\, .
\end{equation}

In all simulations, the initial radio source was modeled as a point source with an isotropic distribution of wavevector $\vec{k}$ and with a frequency $\omega = 1.1 \, \omega_{pe}(R_s)$ corresponding to the near-fundamental plasma emission at a distance $R_s$ from the solar center, determined using a spherically symmetric Parker density model \citep{1960ApJ...132..821P} with constant temperature and constants chosen to agree with satellite measurements adapted from \cite{1999A&A...348..614M}. The absolute value of the wavevector $k = (\omega ^2 - \omega_{pe}^2)^{1/2}/c$ is therefore the same for all photons. Although this density model is relatively simple and has been used successfully for the simulations of Type III bursts in the past \citep[e.g.][]{2001SoPh..202..131K}, it does not have a simple analytical form, which is needed for the solution of the differential equations~(\ref{eq:photon}). To simplify the density model, we fit the numerical solution with three power-law functions \citep[see, e.g.,][]{2018PhDT_ben}, giving

\begin{equation}\label{eq:density}
  n(r)= 4.8\times 10^{9}\left(\frac{R_{\odot}}{r}\right)^{14}
  + 3\times 10^8\left(\frac{R_{\odot}}{r}\right)^6
  +1.4\times 10^6\left(\frac{R_{\odot}}{r}\right)^{2.3} \,\,\, ,
\end{equation}
which can be easily differentiated to find the derivatives useful to solve the ray-tracing Equations~(\ref{eq:photon}).

The simulations begin with approximately $10^4$ photons with different initial positions given by $R_s$ from 1.05 - 57 $R_{\odot}$. Using the coronal density model given by Equation~(\ref{eq:density}), these correspond to plasma frequencies from 460 to 0.1~MHz, respectively. The photon transport was simulated until a distance where both refraction and scattering become negligible, or until the photon frequency $\omega$ (which is conserved in the simulations) became much larger than the local plasma frequency. In each simulation run, a photon was traced until it crossed a sphere where scattering becomes negligible or to 1 AU (whatever is less) and the arrival time and photon properties at this sphere were recorded. The locations of the photons on this sphere directed toward the observer (i.e., those with $0.9<k_z/k<1$) were then back-projected to the source plane, thus defining the apparent source intensity map $I(x,y)$
\citep[][red region in Figure~\ref{fig:cartoon}]{2010A&A...513L...2K}. Similarly, the spread of arrival times on this sphere determines the observed burst intensity time profile. In order to calculate the decay time at different frequencies, we first select the peak time of the flux (maximum in the histogram of the arrival times); times greater than the peak time are regarded as defining the decay phase, which was fitted with a Gaussian form. The delay time is defined as the half width at half maximum (HWHM) of the Gaussian fit.

The total flux was evaluated by performing an integral $\int I(x,y) \, dx \, dy$ over the corresponding source area.  Also, using solar disk-centered coordinates, the centroid position of the source ($\bar{x}$, $\bar{y}$) was found by calculating the first normalized moments (means) of the distribution:

\begin{equation}
\bar{x}=\frac{\int_{-\infty}^{\infty}x \, I(x,y) \, dx \, dy}{\int_{-\infty}^{\infty} I(x,y) \, dx \, dy}\;\;\; \mbox{,} \qquad
\bar{y}=\frac{\int_{-\infty}^{\infty}y \, I(x,y) \, dx \, dy}{\int_{-\infty}^{\infty}I(x,y) \, dx \, dy}
\label{eq:xy_c}
\end{equation}
and the variances ($\sigma^2_x$, $\sigma^2_y$) calculated using the second normalized moments:

\begin{equation}
\label{eq:variances}
\sigma^2_x = \frac{\int_{-\infty}^{\infty}(x-\bar{x})^2 \, I(x,y) \, dx \, dy}{\int_{-\infty}^{\infty} I(x,y) \, dx \, dy}
\; \mbox{,} \qquad
\sigma^2_y = \frac{\int_{-\infty}^{\infty}(y-\bar{y})^2 I(x,y) \, dx \, dy}{\int_{-\infty}^{\infty} I(x,y) \, dx \, dy} \,\,\, .
\end{equation}
The FWHM in each direction can then be calculated using

\begin{equation}\label{eq:FWHMxy}
  {\rm FWHM}_{x,y} = 2\sqrt{2\ln{2}} \, \sigma_{x,y},
\end{equation}
based on the assumption that the distribution $I(x,y)$ is Gaussian. To evaluate the FWHM source sizes we also fitted $I(x,y)$ with a 2D Gaussian and determined the sizes using the best-fit parameters. Typical images $I(x,y)$ are shown in Figures~\ref{fig:eps08_anis05} and~\ref{fig:eps08_anis03}.

Because of the finite number of photons in the sample, the source centroids (Equation~(\ref{eq:xy_c})) and sizes (Equations~(\ref{eq:variances})) have associated statistical errors \citep[see, e.g.,][]{Rao1973}. The uncertainties in the mean values can be estimated as

\begin{equation}\label{eq:dcentroid}
  \delta \bar{x}\simeq \frac{\sigma _x}{\sqrt{N}}\,,
  \;\;\;\;\;
  \delta \bar{y}\simeq  \frac{\sigma _y}{\sqrt{N}}\,,
\end{equation}
and the uncertainty in the FWHM sizes as

\begin{equation}\label{eq:dFWHM}
  \delta \, {\rm FWHM}_{x,y} \simeq 2\sqrt{2\ln{2}} \, \frac{\sigma_{x,y}}{\sqrt{2N}} \,\,\, ,
\end{equation}
where $N\gg1$ is the number of photons used to determine the means $(\bar{x},\bar{y})$ and the standard deviations $\sigma_x, \sigma_y$. These uncertainties are used in all numerical results presented in this paper.

\citet{2018ApJ...857...82K} have recently investigated the effects of isotropic scattering on time-profiles generated in the interplanetary medium using Monte Carlo simulations. They assumed a power-law spectrum of electron density fluctuations (see Appendix \ref{app:pl} for the derivation) and also used expressions for the diffusion coefficient from \cite{2007ApJ...671..894T} and \cite{2008ApJ...676.1338T} to describe the scattering effects. We adopt the same density fluctuations model here. \cite{2018ApJ...857...82K} used\footnote{Note a missing factor of $\pi/2$ in their equation.} Equation (\ref{eq:q_pl_spectr-5-3}), viz.

\begin{equation}\label{eq:q_krupar}
  \bar{q} \, \eps ^2 \simeq 4\pi l_0^{-2/3}l_i^{-1/3} \eps ^2 \,,
\end{equation}
where $l_i=(r/R_\odot)$~[km] is the inner scale of the electron density fluctuations \citep{1987sowi.conf...55M, 1989ApJ...337.1023C}, $R$ is the heliocentric distance, $l_o=0.25R_\odot (R/R_\odot)^{0.82}$ is an empirical formula for the outer scale \citep{2001SSRv...97....9W}, and $\epsilon = \sqrt{\langle \delta n^2 \rangle/n^2}$ is the level of density fluctuations with the spectrum given by Equation (\ref{eq:pl_spectr}). $\eps$ was taken as a quantity independent of radial distance.

We stress that for the density fluctuations spectrum (\ref{eq:pl_spectr}), the scattering rate is determined by the density fluctuations at scales near $l_i$. Since both the density fluctuations variance $\eps ^2$ and the outer scale $l_o(r)$ determine the level of density fluctuations in Equations~(\ref{eq:q_krupar}), $\eps (r)$ cannot be determined without knowledge of $l_0(r)$, and different models for $l_0(r)$ result in different values for $\eps (r)$.  Hence the $\eps $ values taken for the simulations in the next section should be viewed as the standard deviation of density fluctuations for a given outer scale model $l_o(r)$, and may not be suitable for direct comparison with density fluctuation measurements in the corona.

\subsection{Simulation results for a single frequency}

\begin{figure}
\centering
\includegraphics[trim=0.5cm 0.1cm .9cm .1cm, clip, width=0.329\linewidth]{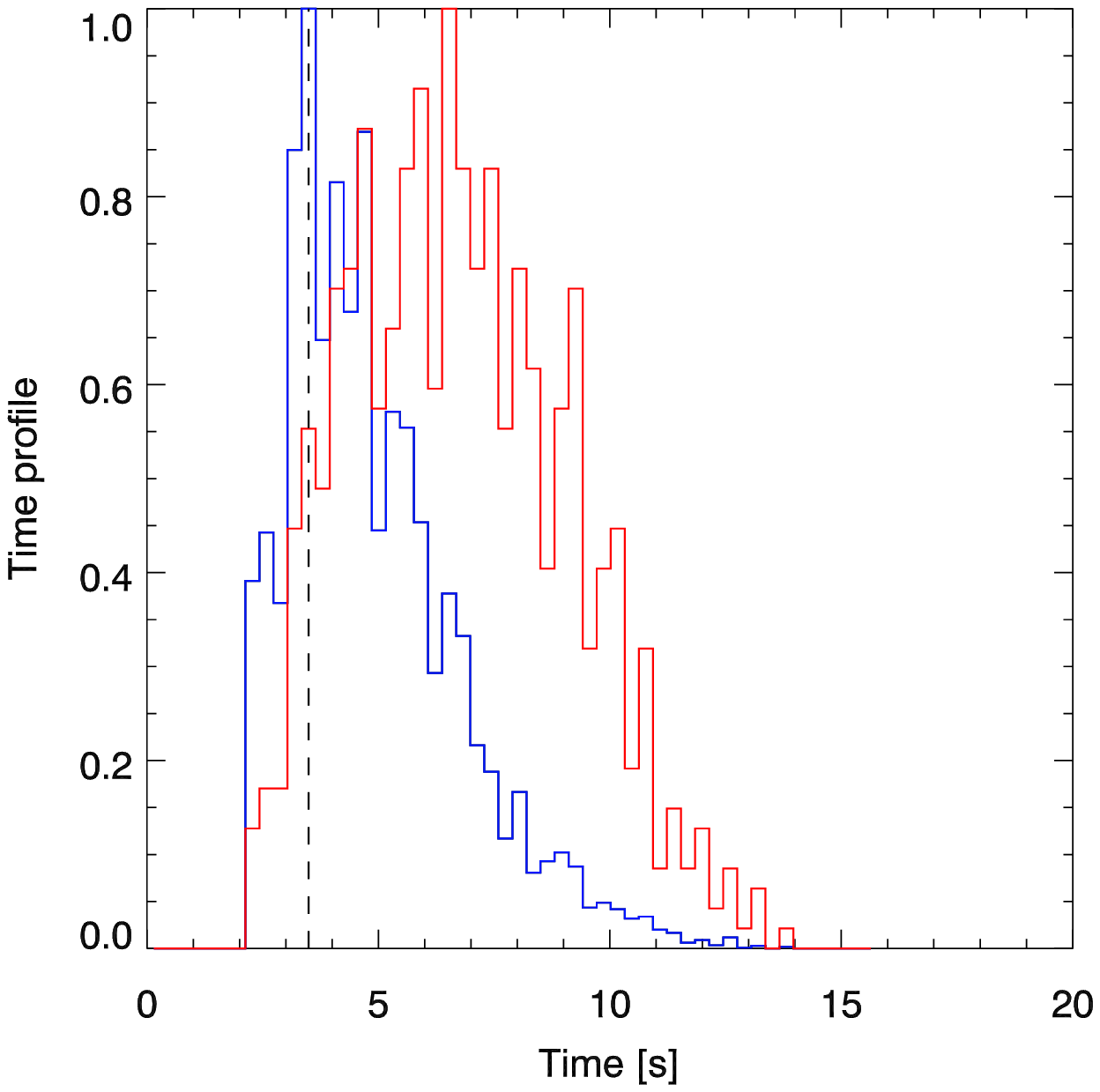}
\includegraphics[trim=0.5cm 0.1cm .9cm .1cm, clip, width=0.329\linewidth]{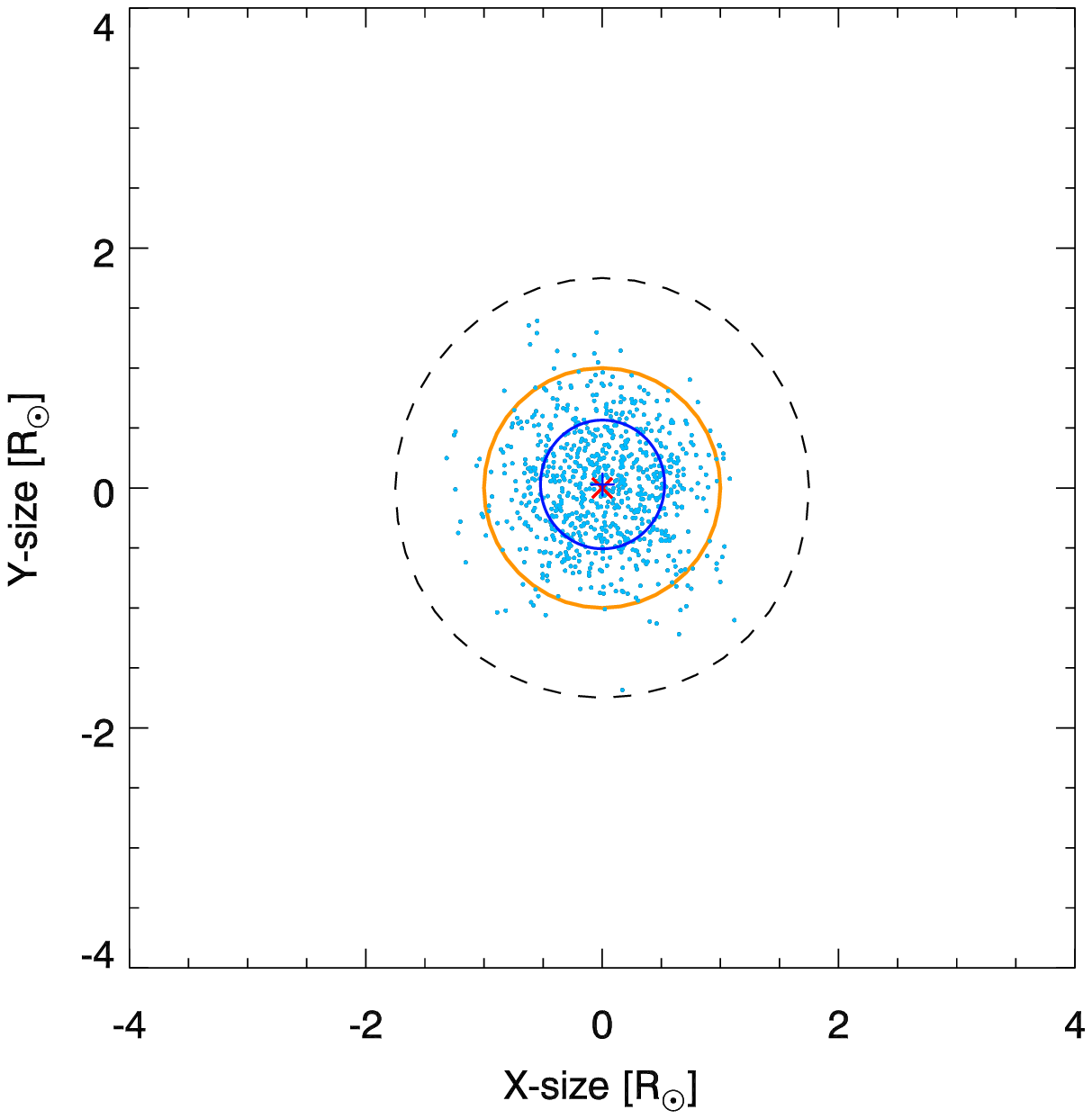}
\includegraphics[trim=0.5cm 0.1cm .9cm .1cm, clip, width=0.329\linewidth]{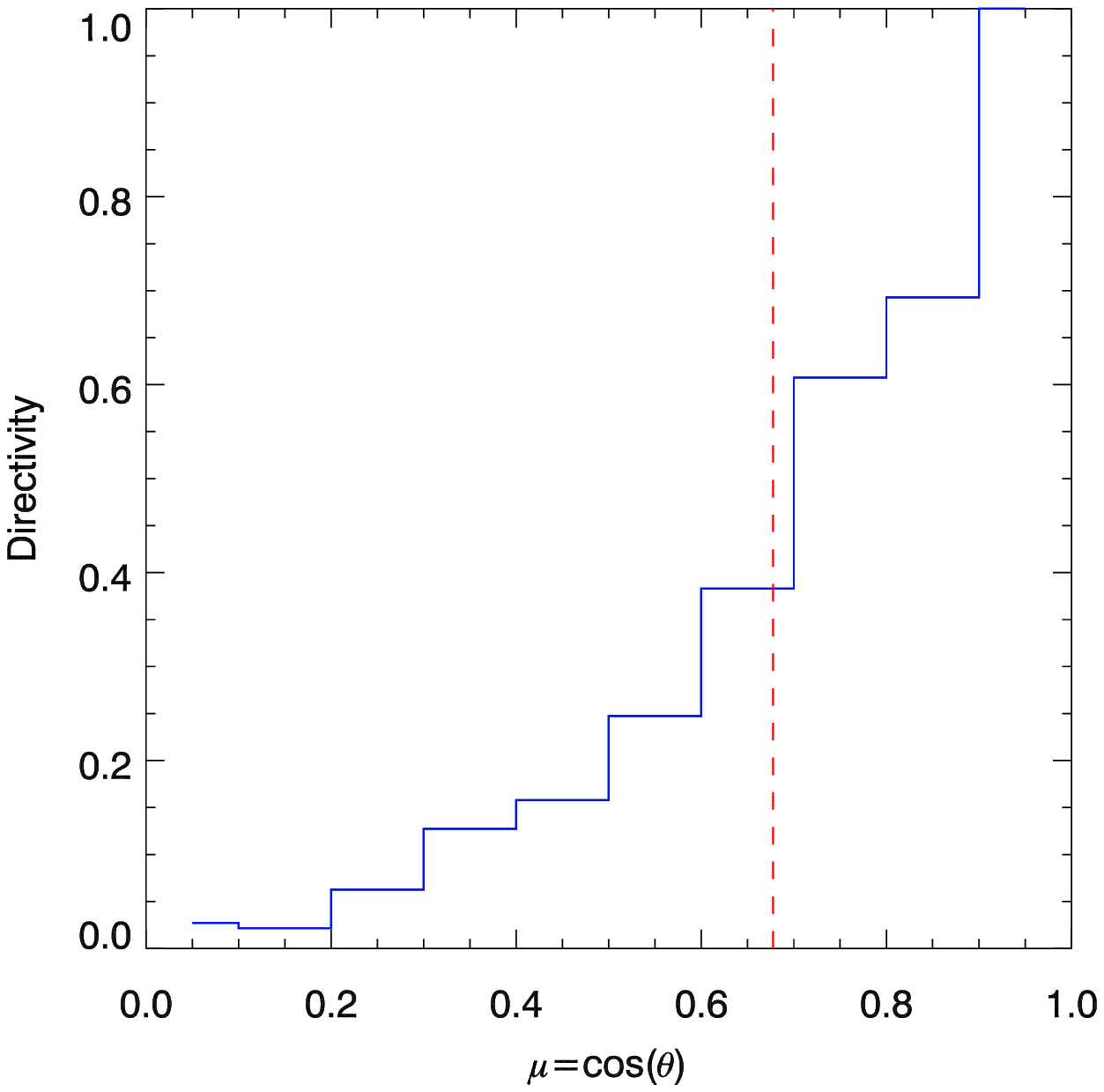}
\caption{\label{fig:eps08_anis05} Simulations for a point source located at  $R_S=1.75R_\odot$ ($f_{pe}=32$~MHz), and using $\eps=0.8$, $\alpha=0.5$.
\textit{Left:} Time profile of the observed photons: blue with absorption, red without absorption, dashed line indicates the location of the time-profile maximum;
\textit{Center:} Observed radio image in Sun-centered coordinates. The orange circle denotes the Sun, the dashed line denotes the radius where the plasma frequency is 32~MHz, and the blue circle is the FWHM source size. \textit{Right:} Directivity of the observed radio emission. The red dashed line shows the width at half maximum.}
\includegraphics[trim=0.5cm 0.1cm .8cm .1cm, clip, width=0.329\linewidth]{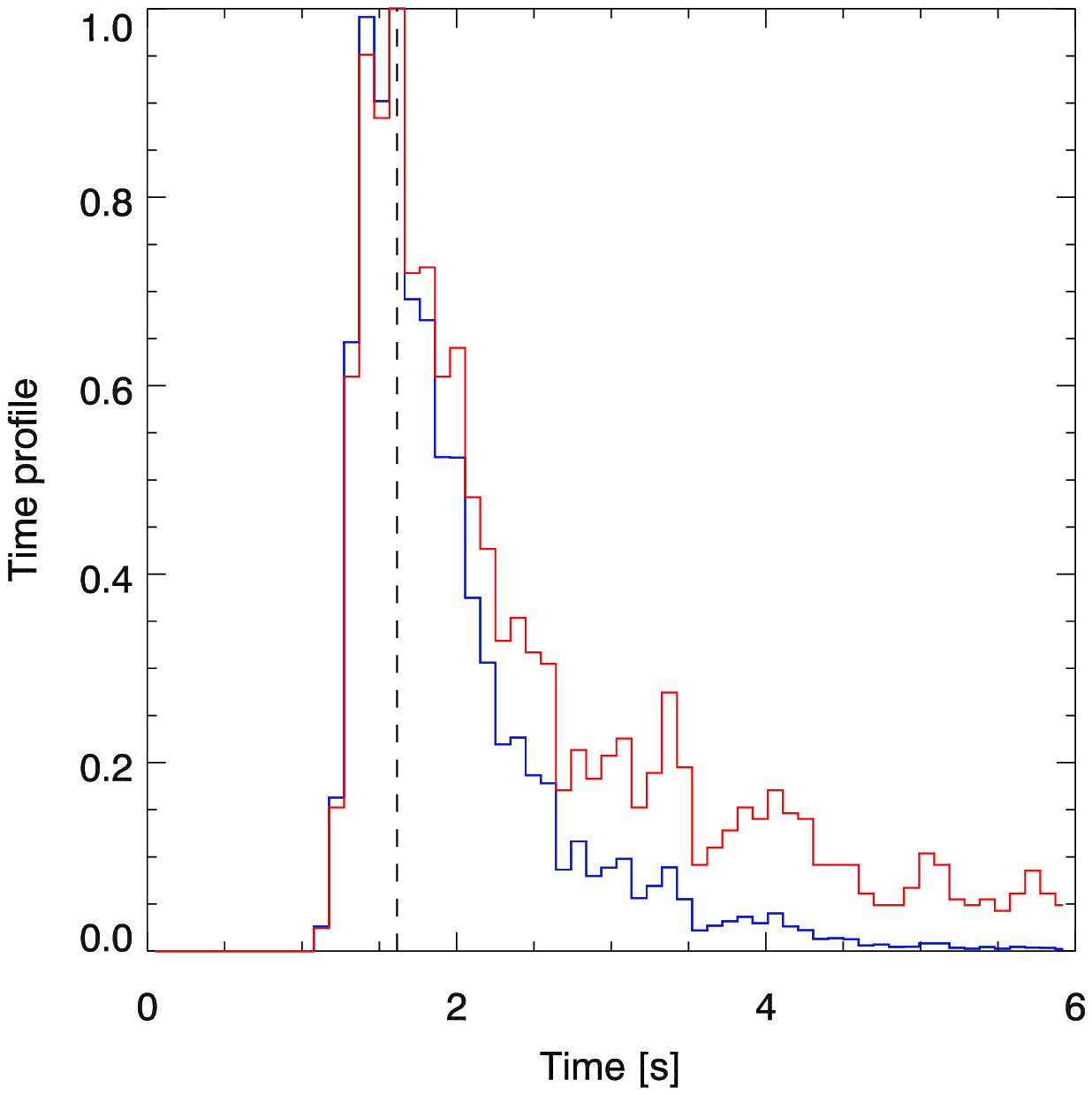}
\includegraphics[trim=0.5cm 0.1cm .8cm .1cm, clip, width=0.329\linewidth]{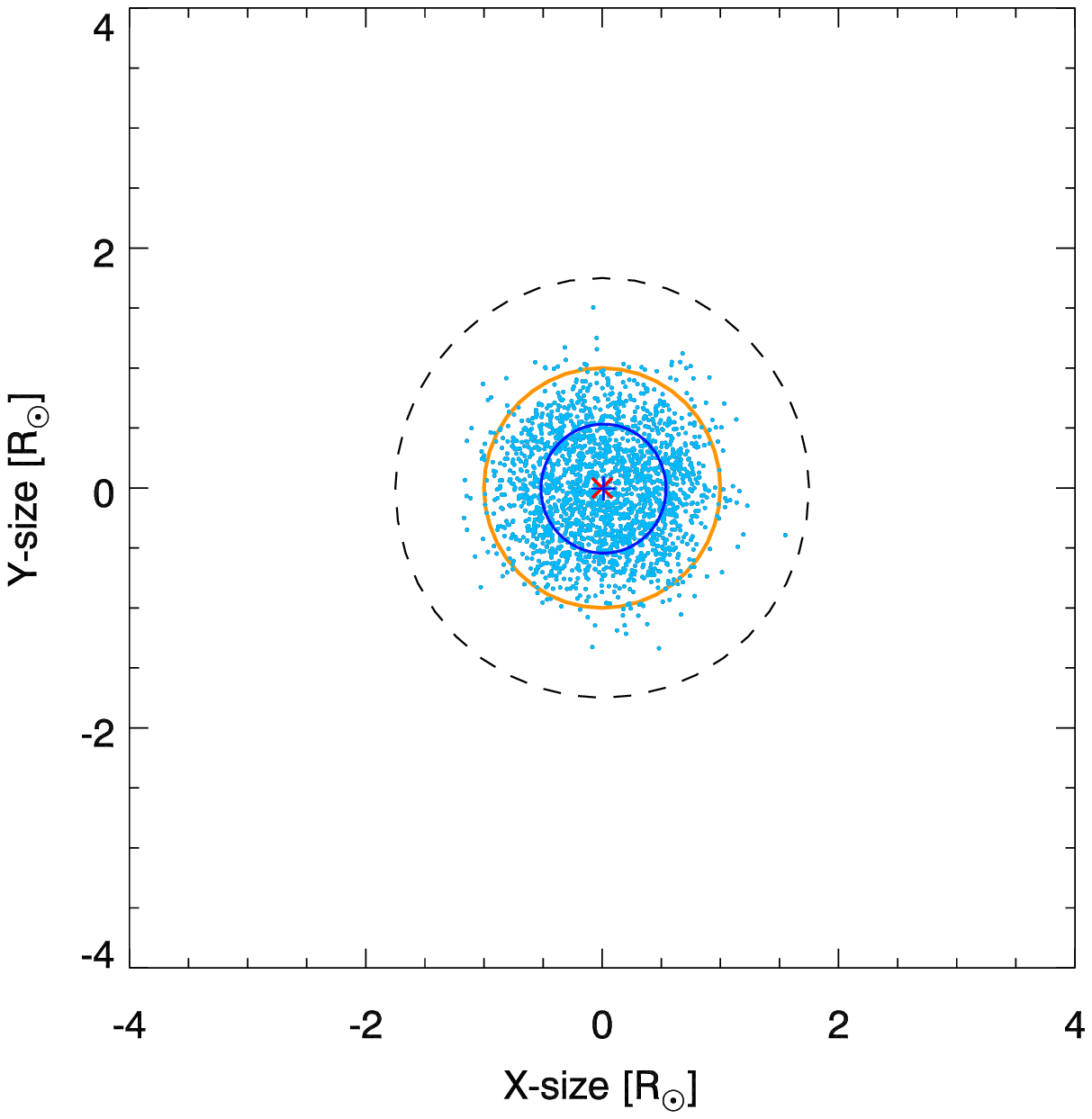}
\includegraphics[trim=0.5cm 0.1cm .8cm .1cm, clip, width=0.329\linewidth]{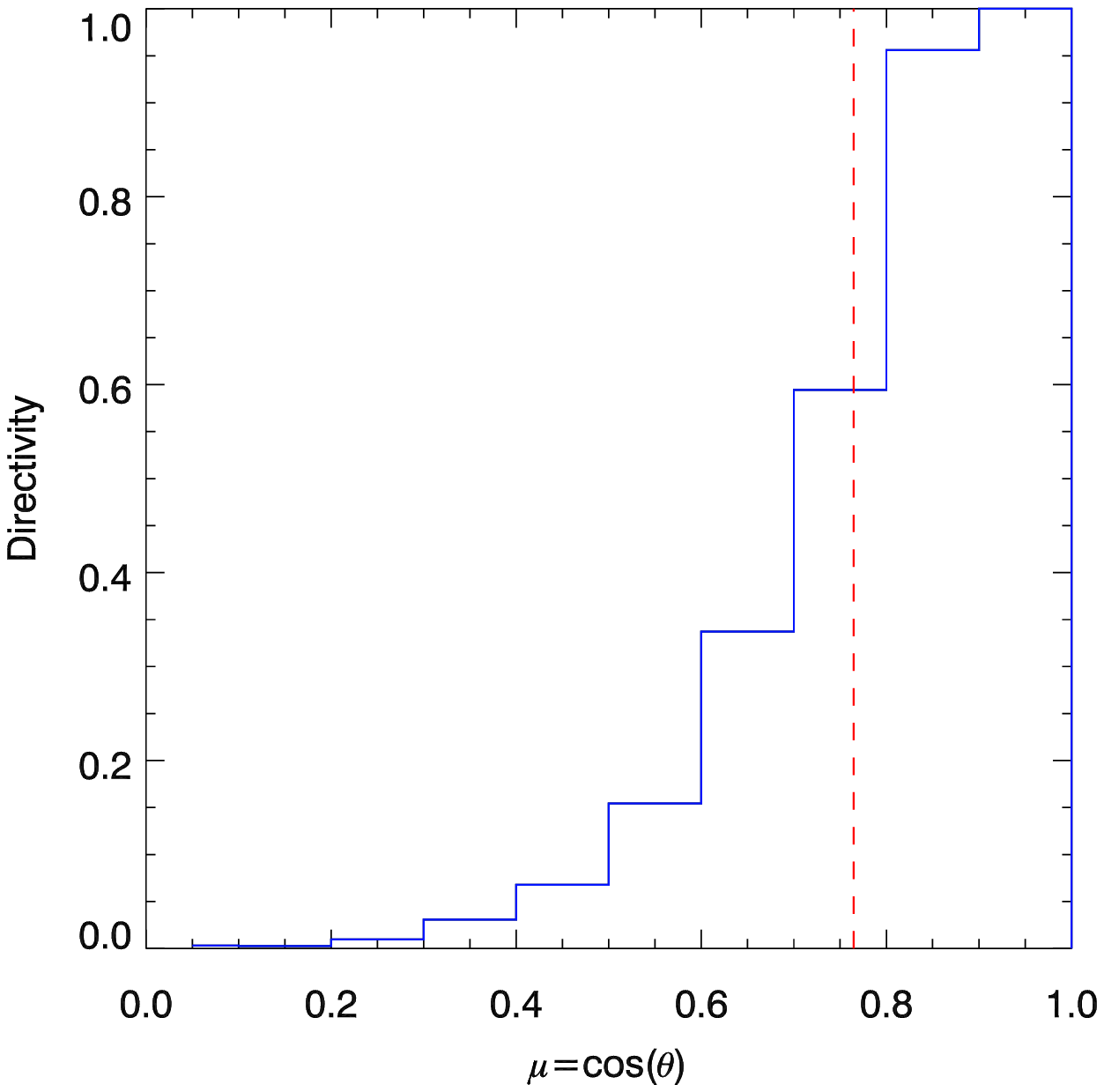}
\caption{\label{fig:eps08_anis03}
Simulation results as in Figure \ref{fig:eps08_anis05} but with stronger anisotropy, $\alpha=0.3$.}
\end{figure}

\begin{figure}
\centering
\includegraphics[width=0.49\linewidth]{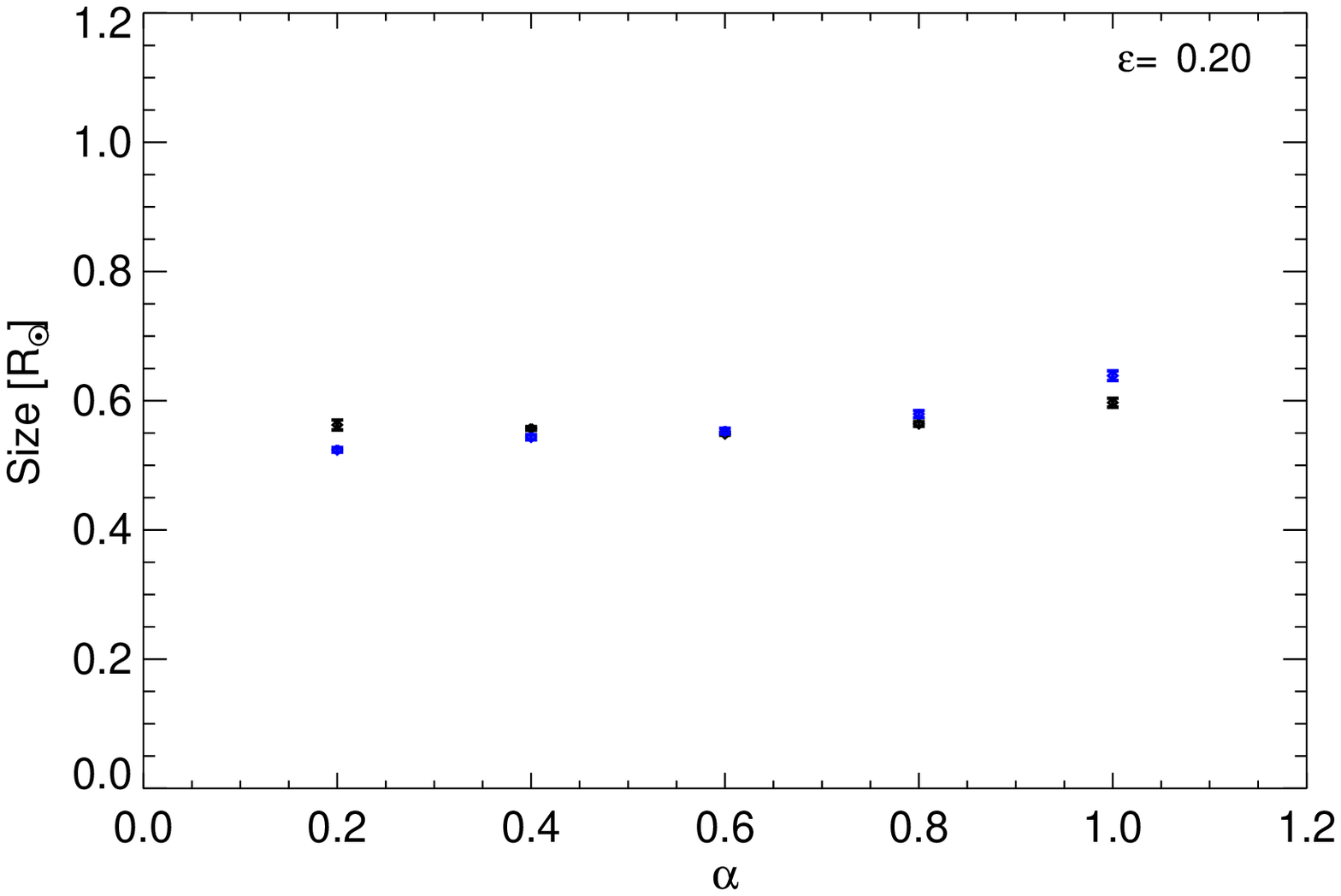}
\includegraphics[width=0.49\linewidth]{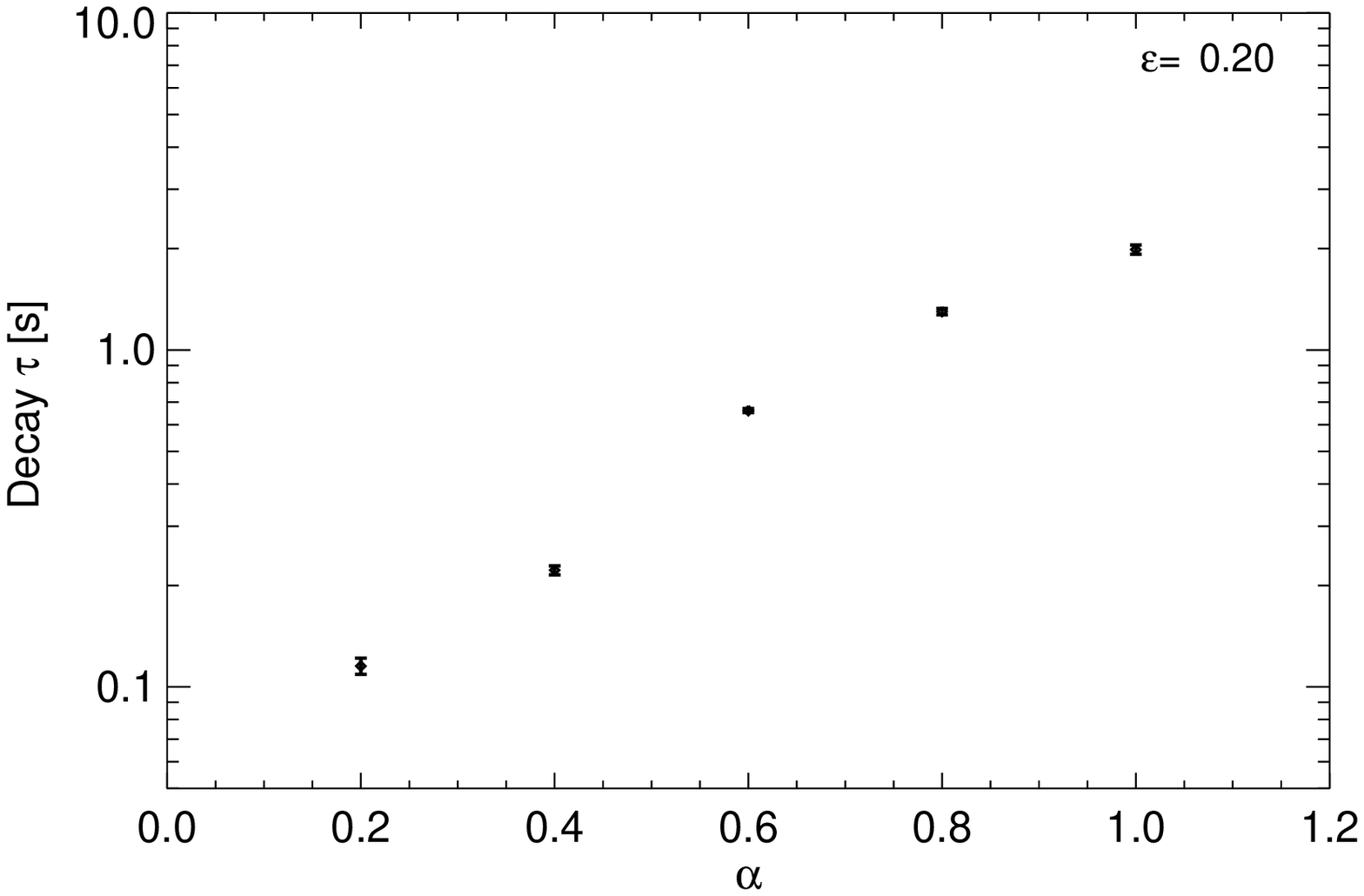}
\caption{\label{fig:size_decay_eps02} FWHM sizes and decay time (HWHM) with $\eps =0.2$ as a function of anisotropy $\alpha$. The black symbols are from fitting the simulation data with a 2D Gaussian function to determine the size and centroid position, the blue sizes are using Equations (\ref{eq:FWHMxy}). One standard deviation uncertainty are calculated using Equation (\ref{eq:dFWHM}). }

\includegraphics[width=0.49\linewidth]{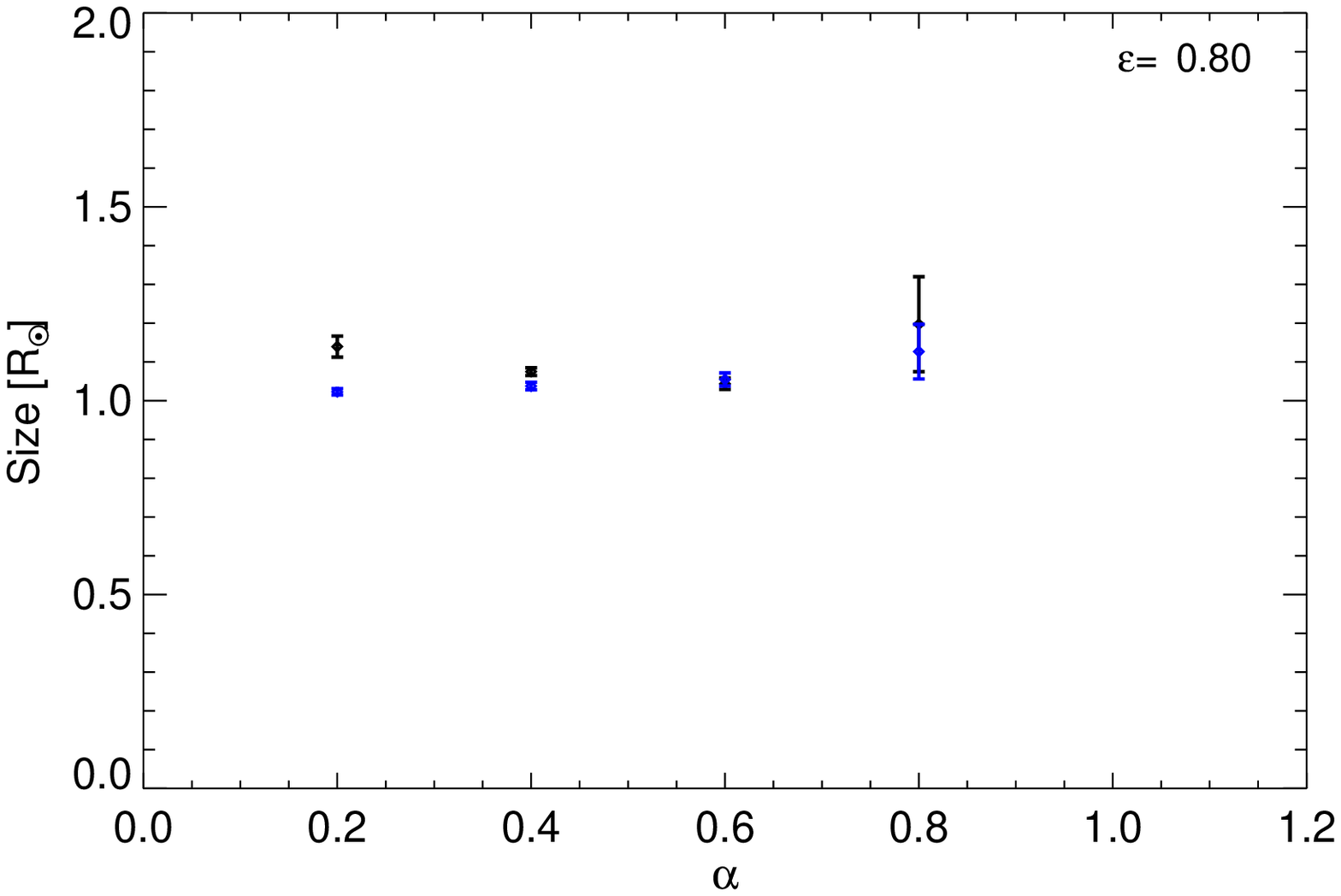}
\includegraphics[width=0.49\linewidth]{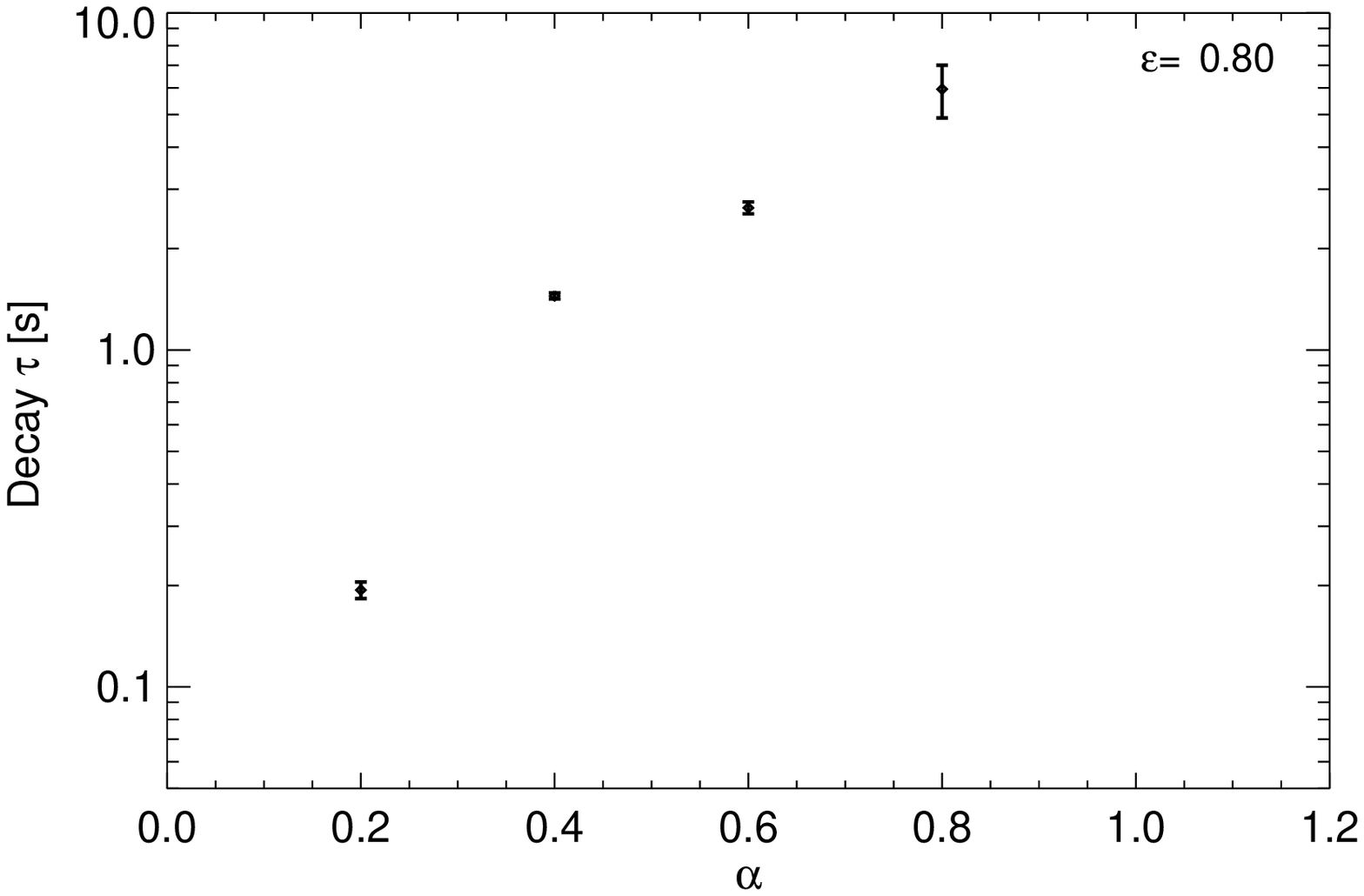}
\caption{\label{fig:size_decay_eps08} The same as Figure \ref{fig:size_decay_eps02}, but for $\eps =0.8$.}
\end{figure}

\begin{figure}
    \centering
\includegraphics[trim=0.5cm 0.1cm .9cm .1cm, clip, width=0.329\textwidth]{obs_source_fpe32_04MHz_FE35_24MHz_eps0_800_anis0_30_ang0_00deg.eps}
\includegraphics[trim=0.5cm 0.1cm .9cm .1cm, clip, width=0.329\textwidth]{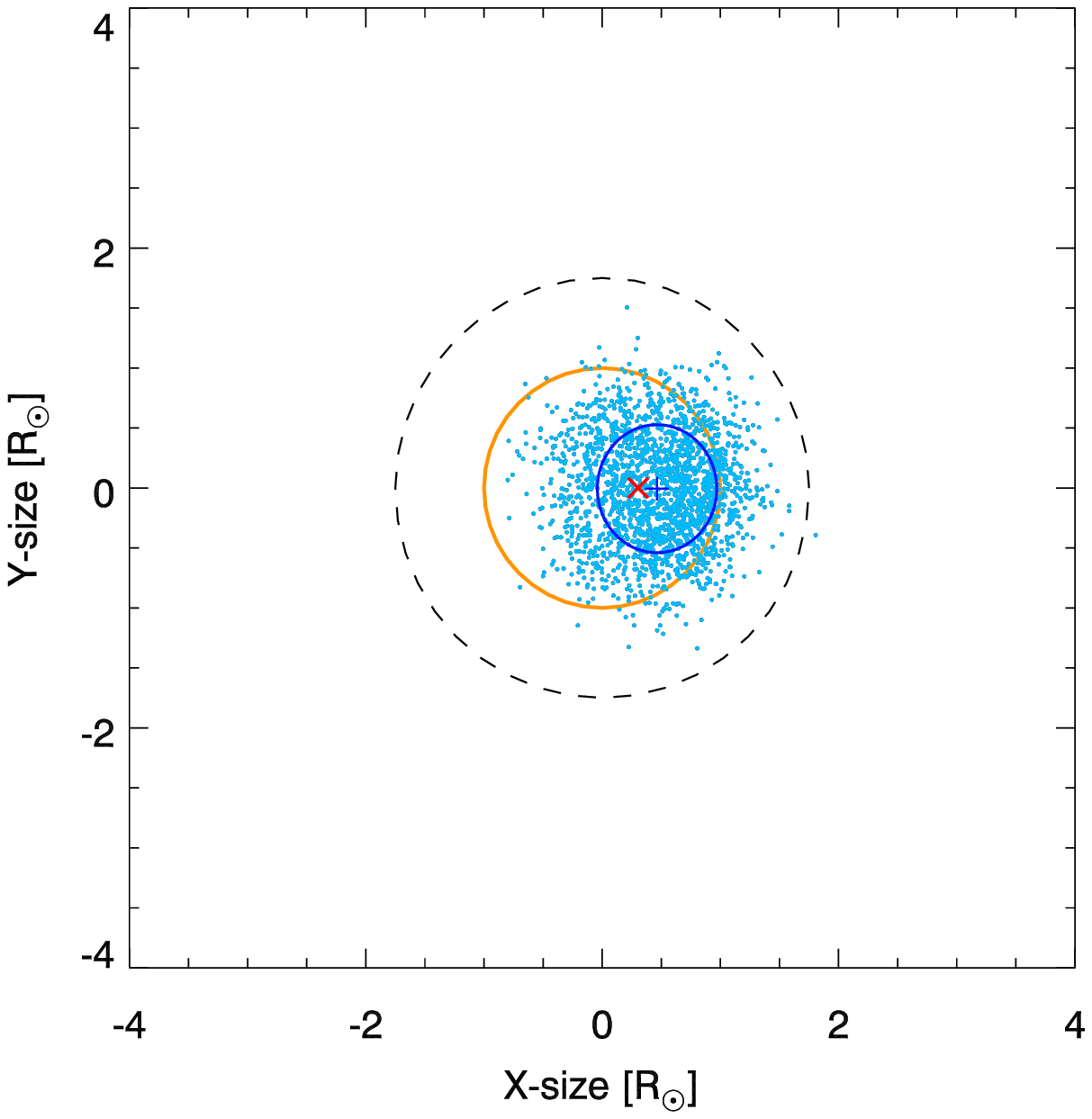}
\includegraphics[trim=0.5cm 0.1cm .9cm .1cm, clip, width=0.329\textwidth]{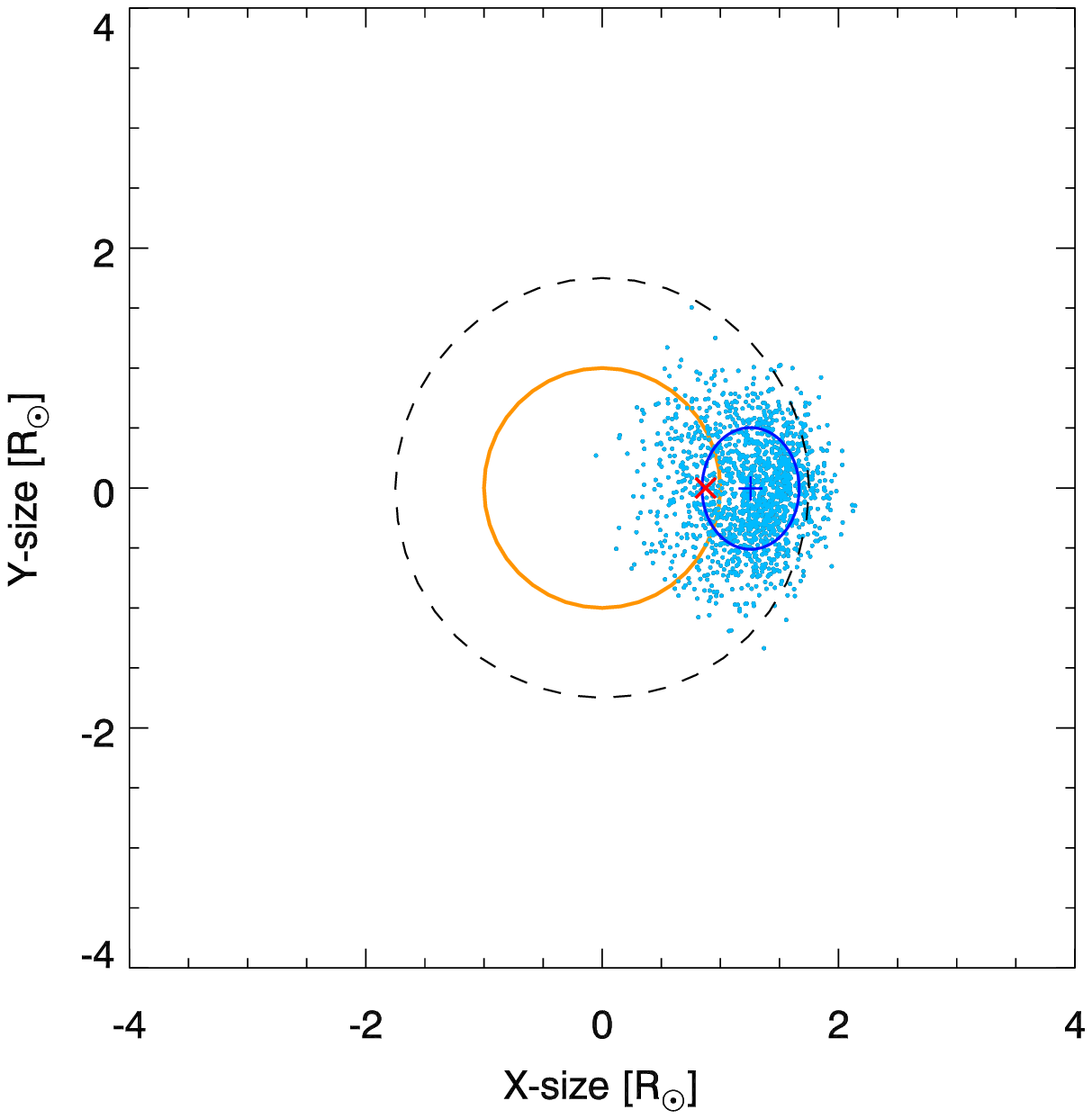}
    \caption{\label{fig:images}  Radio images for a point source located at  $R_S=1.75R_\odot$ ($f_{pe}=32$~MHz), and for three different source locations $\theta_s = 0^o,10^o, 30^o$
    from the disk center. All images are for anisotropic turbulence with anisotropy paramater $\alpha= 0.3$ and a level of turbulence $\eps =0.8$. The projected positions of the source and the image centroid are shown by red and blue crosses respectively. The orange circle denotes the Sun, the dashed line denotes the radius where the plasma frequency is 32~MHz, and the blue circle is the FWHM source size.}
\end{figure}

\begin{figure}
   \centering
\includegraphics[width=0.329\textwidth]{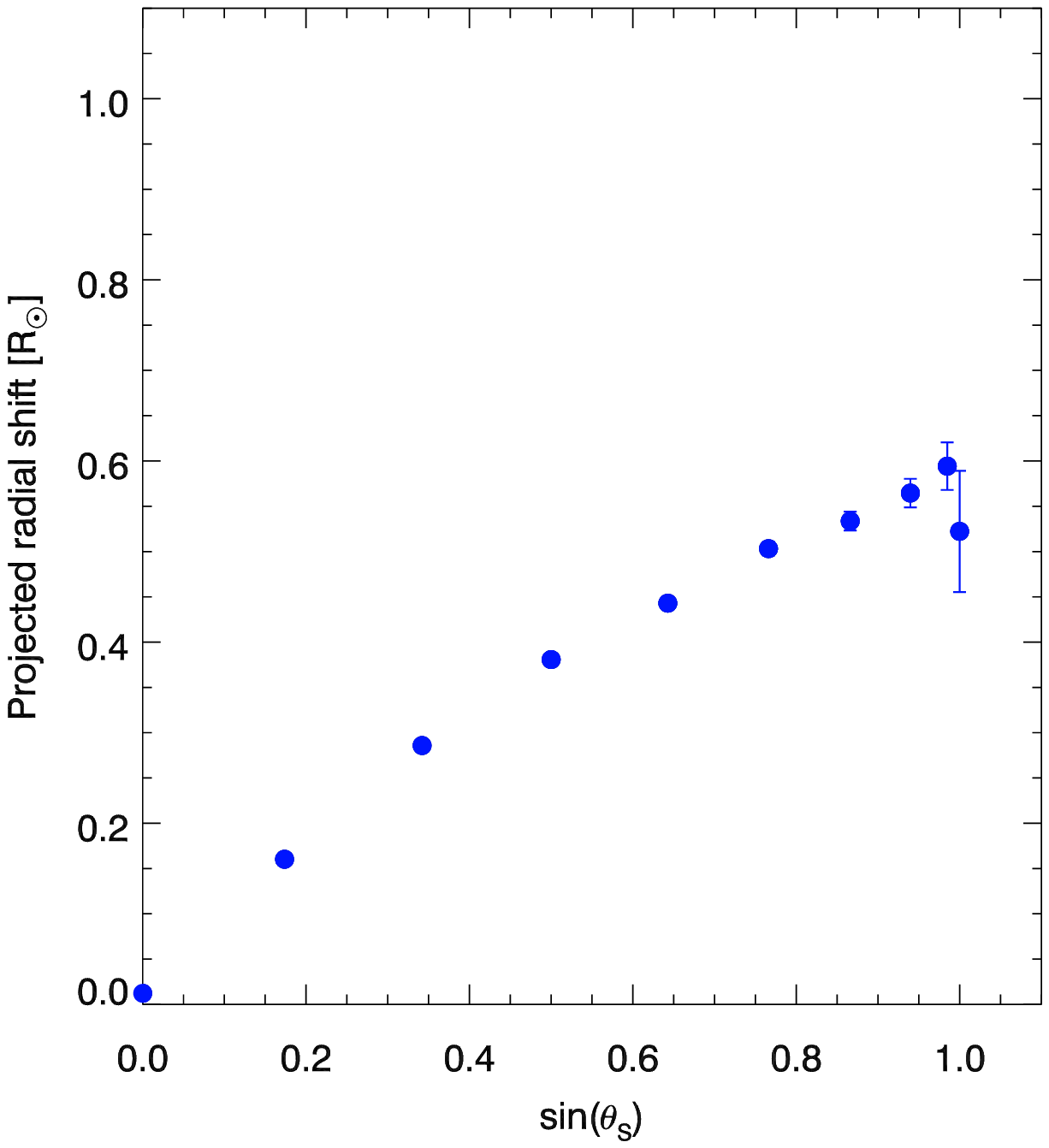}
\includegraphics[width=0.329\textwidth]{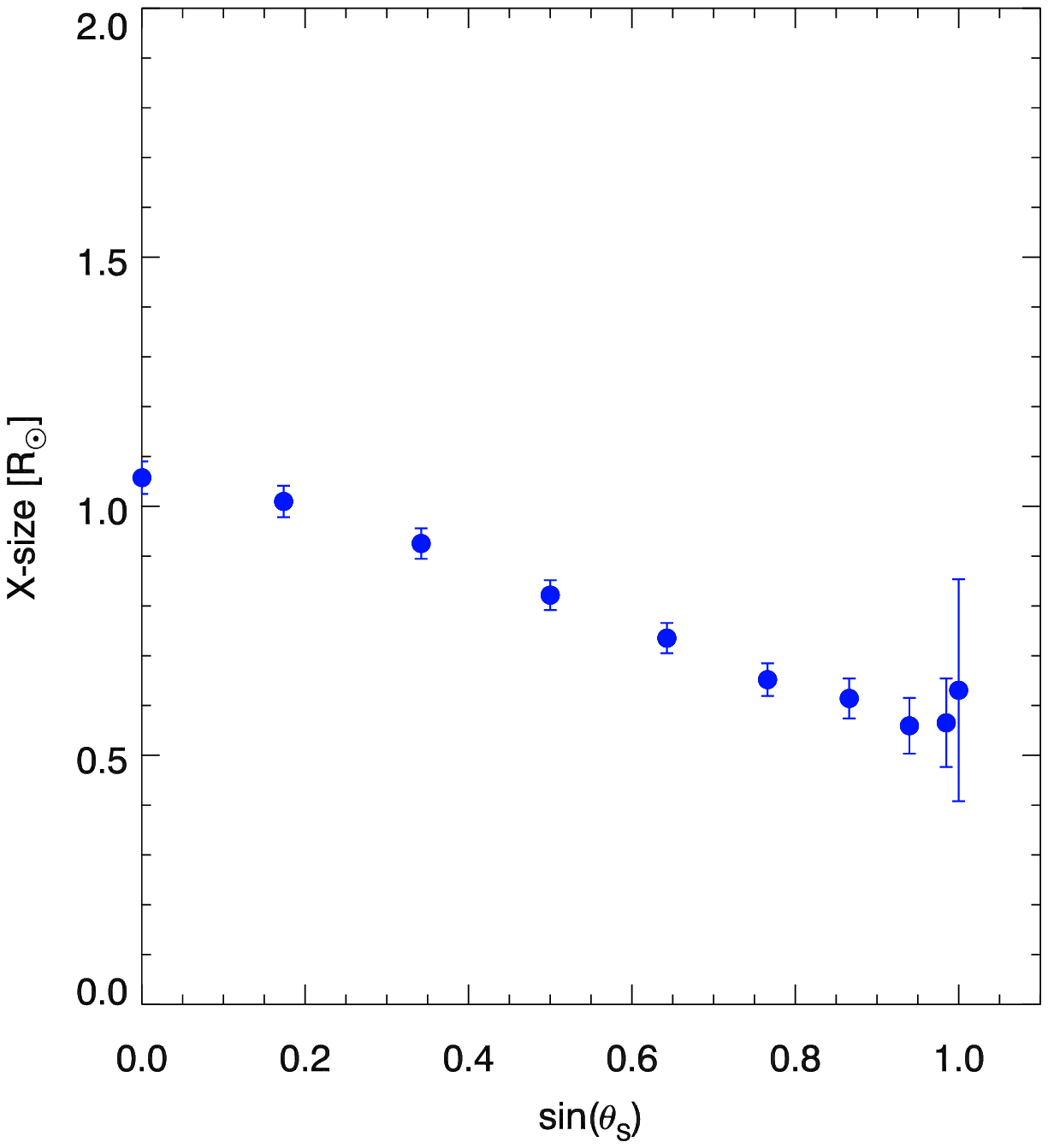}
\includegraphics[width=0.329\textwidth]{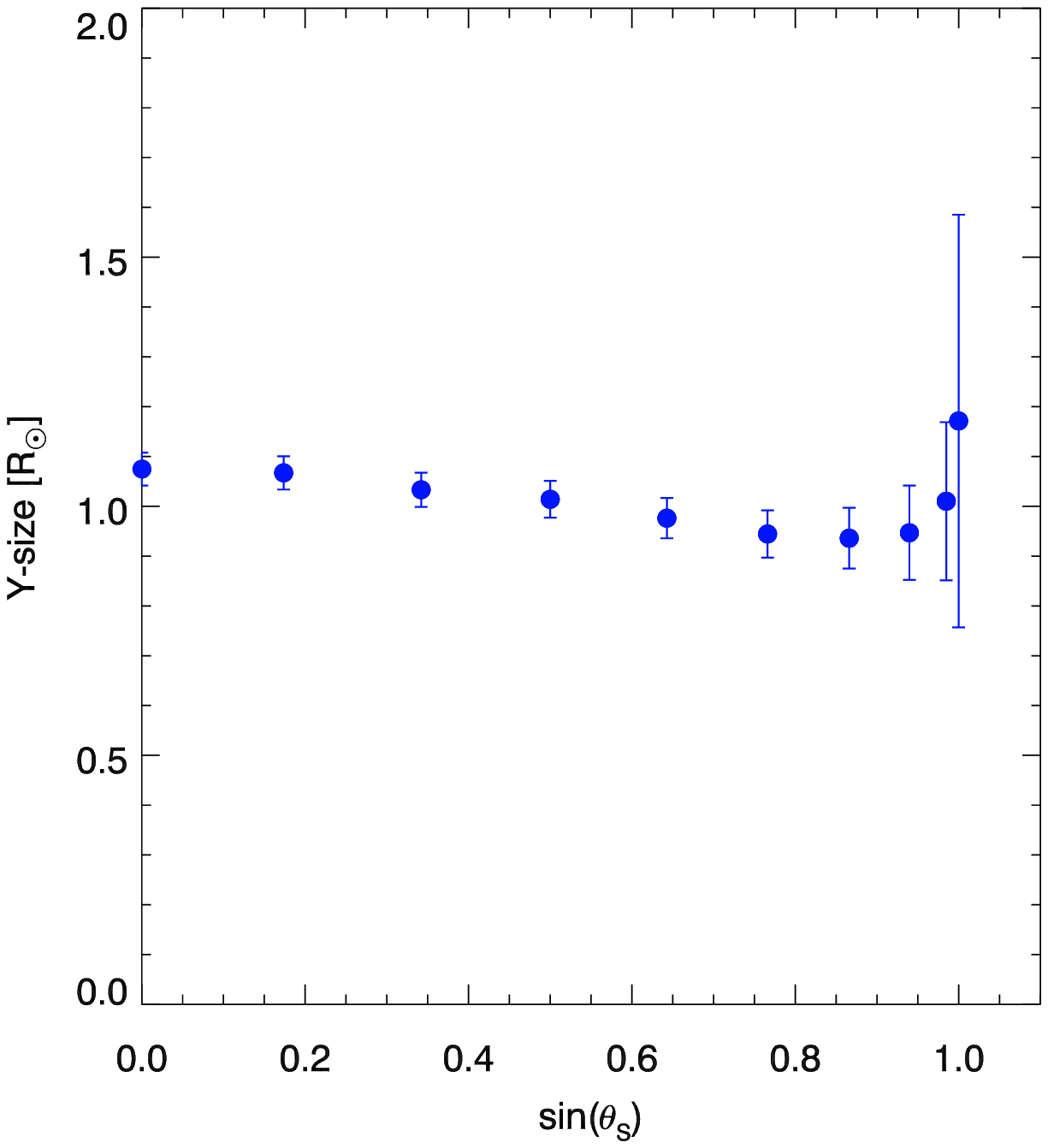}
    \caption{\label{fig:rshift}  \textit{Left:} Shift of the centroid position $\bar{x}$ as a function of the source heliocentric angle $\theta _s$. The shifts are calculated for anisotropic scattering with $\alpha =0.3$ and turbulence level $\eps =0.8$ as in Figures \ref{fig:images}.
     \textit{Center:} FWHM X-size given by Equation~(\ref{eq:FWHMxy}); \textit{right:} FWHM Y-size given by Equation~(\ref{eq:FWHMxy}).
     The error bars show one standard deviations given by Equations (\ref{eq:dcentroid}) and (\ref{eq:dFWHM}). The number of detected photons in $\vec{z}$-direction is decreasing, so the uncertainties are large for angles close to $\theta _s\simeq 90^o$.}
\includegraphics[width=0.329\textwidth]{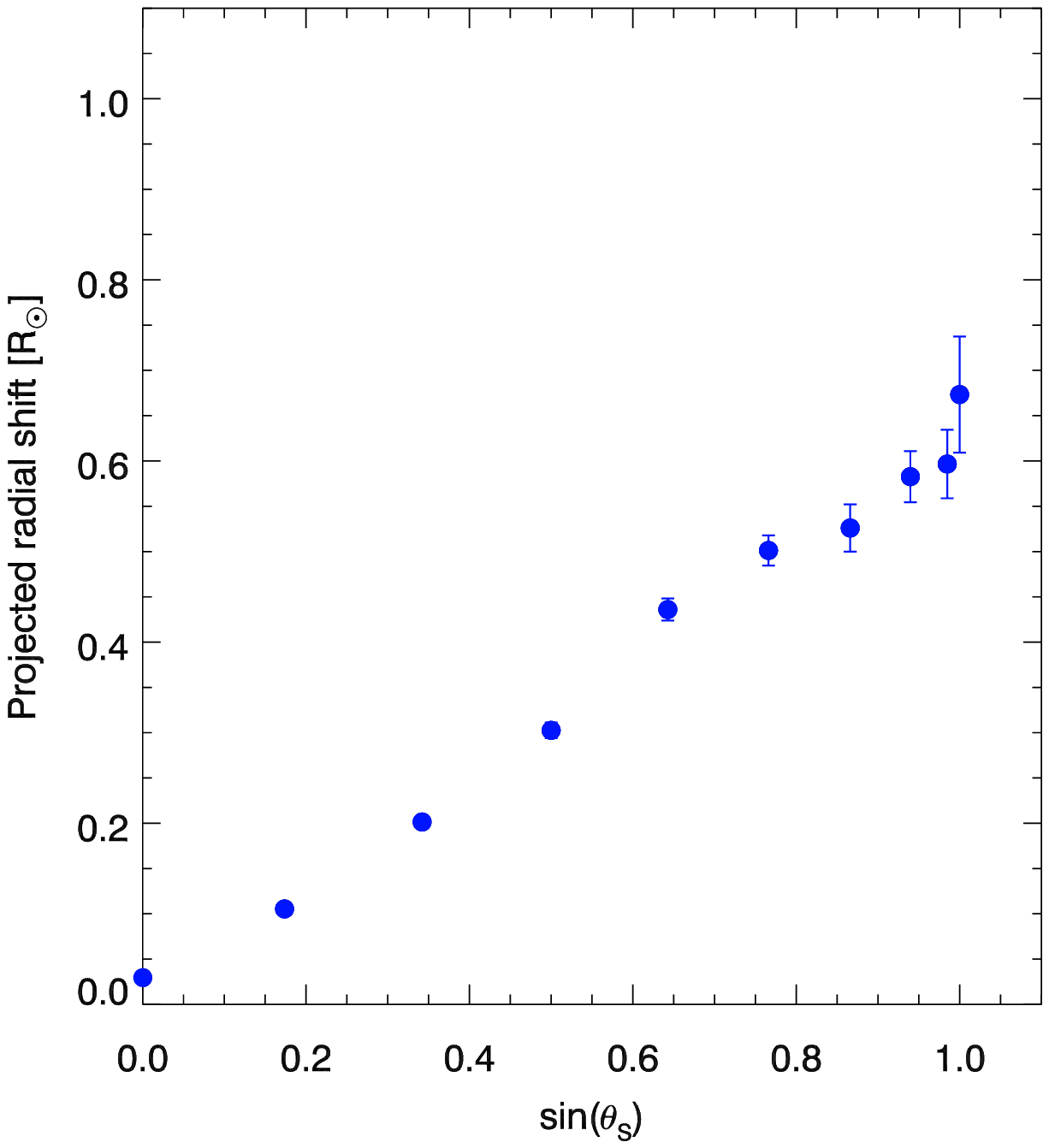}
\includegraphics[width=0.329\textwidth]{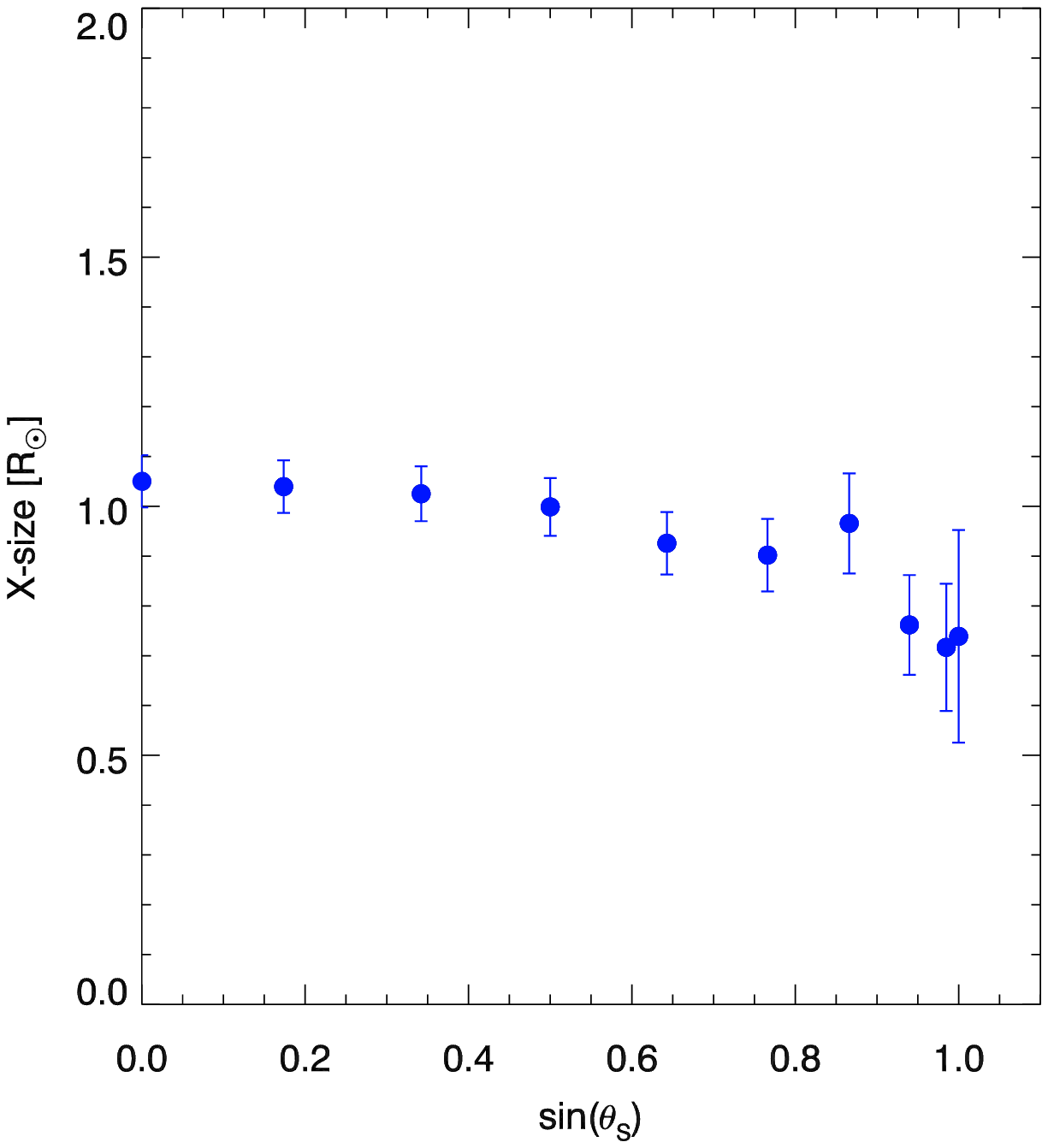}
\includegraphics[width=0.329\textwidth]{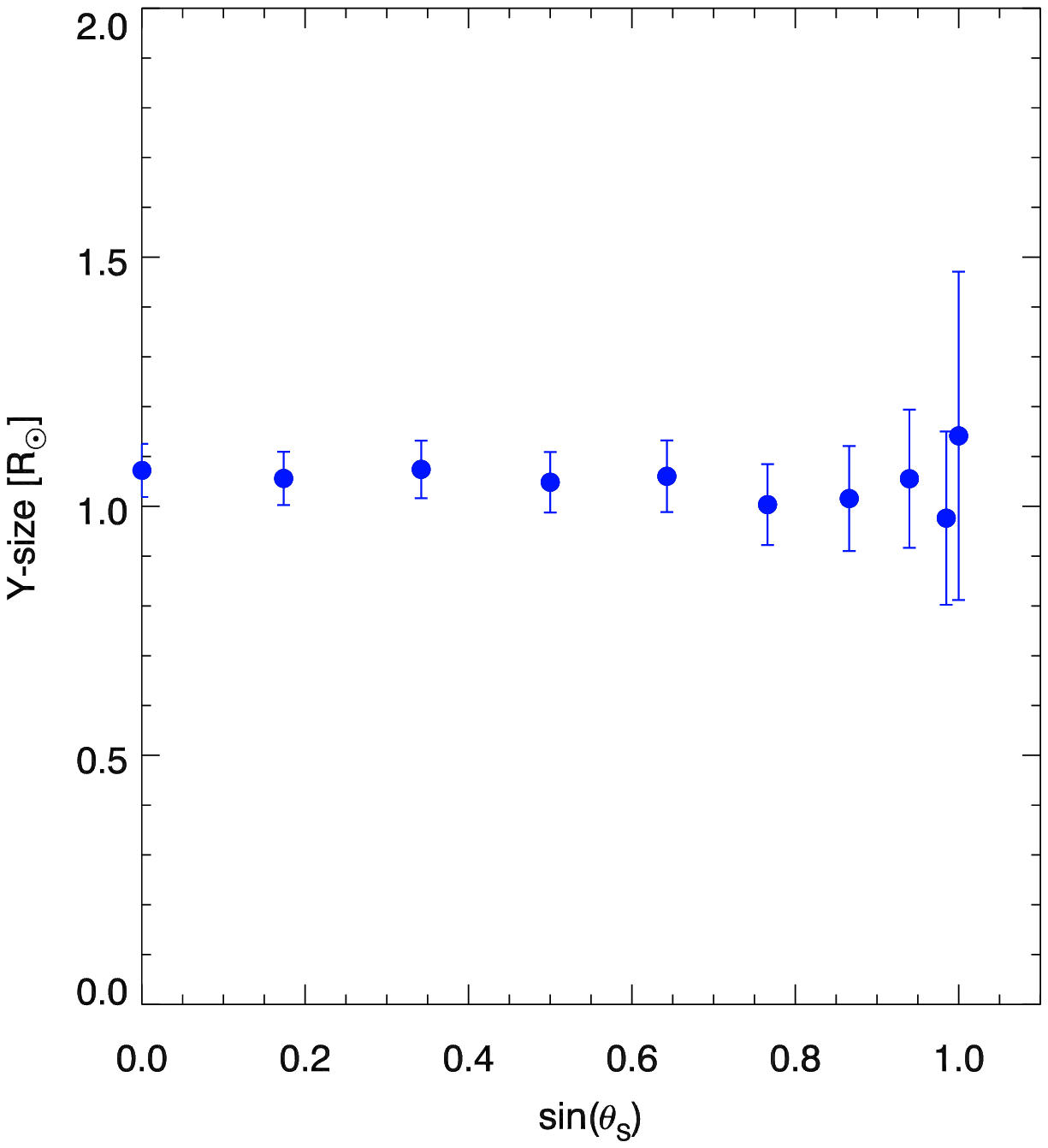}
    \caption{\label{fig:rshift05}  The same as Figure \ref{fig:rshift},
    but for $\alpha=0.5$. }
\end{figure}

Using the assumptions presented in the previous section, we can choose $\eps$ so that the characteristic size of the radio source is about $19'$ for $f_{pe}=32$~MHz (observing frequency $\sim 35$~MHz), as typically observed for fundamental plasma emission \citep{2017NatCo...8.1515K}. Figures~\ref{fig:eps08_anis05} to~\ref{fig:rshift} plot the main results of the ray-tracing simulations. Figures~\ref{fig:eps08_anis05} and~\ref{fig:eps08_anis03} show the results for a point source located above the solar disk center at a height $0.75R_\odot$ above the photosphere, where $f_{pe}=32$~MHz according to the density model~(\ref{eq:density}). The simulations presented in Figures~\ref{fig:eps08_anis05} use the same level of density fluctuations $\eps$ but different values of the anisotropy parameter ($\alpha=0.3$ and $\alpha=0.5$, respectively). For both cases, the FWHM source size is about $1.15R_\odot$ (consistent with $19'$ FWHM size observations), but the time profile for the simulation with $\alpha = 0.5$ (Figure~\ref{fig:eps08_anis05}) is significantly broader than that for $\alpha = 0.3$ (Figure~\ref{fig:eps08_anis03}). Turbulent density fluctuations which have a power that is weaker in the parallel direction compared to the perpendicular to radial direction result in a reduced time-broadening effect (i.e. radio-wave cloud broadening along the $z$ direction); consequently the results with anisotropy factor $\alpha = 0.3$ give a characteristic decay time $\sim 0.6$~s, exactly as observed \citep[see Figure~4 in][]{2018SoPh..293..115S}.

Figures~\ref{fig:size_decay_eps02} and~\ref{fig:size_decay_eps08} demonstrate how the observed source sizes and the decay times vary with the value of the anisotropy parameter $\alpha$. Low-level density fluctuations (e.g., $\eps =0.2$; Figure~\ref{fig:size_decay_eps02}) are too weak to provide sufficient scattering to explain FWHM sizes as large as $1.15R_\odot$. At the same time, nearly isotropic scattering (Figure \ref{fig:size_decay_eps08}) with $\eps =0.8$ provides the observed sizes, but the decay time appears to be larger than observed. Reduced scattering along the radial direction (e.g., density fluctuations that are predominantly in perpendicular directions) decreases the characteristic decay time and anisotropy, and a value $\alpha =0.3$ provides the best match to the observations. Indeed, comparing Figures~\ref{fig:size_decay_eps02} and~\ref{fig:size_decay_eps08}, we find that a density fluctuation level of $\eps \simeq 0.8$ and an anisotropy parameter of $\alpha =0.3$ are the parameters that best explain recent LOFAR observations by \citet{2017NatCo...8.1515K} and consistent with the source sizes reported by \citet{1980A&A....88..203D}.

Scattering of photons close to the intrinsic source contributes substantially to the free-free absorption of radio waves. Photons experiencing strong scattering stay longer in the collisional medium and hence are absorbed. Indeed, Figure~\ref{fig:eps08_anis05} demonstrates that the time profile is significantly extended when absorption is switched off. This difference is smaller for the stronger anisotropy case presented in Figure~\ref{fig:eps08_anis05}.

The main effects on source location and size are shown in Figure~\ref{fig:images}.  Because of projection effects along the radial direction, the FWHM source size along the $x$-direction decreases with heliocentric angle (Figure \ref{fig:rshift}), while the FWHM in the $y$-direction (perpendicular to the radial direction) changes only weakly, remaining $1-1.2R_\odot$.
Sources located away from the disk center are shifted radially (along $x$-direction in our simulations), and the near-linear dependence of the source position on $\sin\theta_s$ can be clearly seen from Figure \ref{fig:rshift}. The observer sees an apparent position that is shifted radially away from the disk center, with the shift projected onto the skyplane proportional to $\sin\theta _s$. While sources near the disk center $\theta _s =0$ are radially shifted towards the observer, the true and apparent sources coincide in the $(x,y)$ plane of the sky. The case of more isotropic scattering (Figure \ref{fig:rshift05}) suggests that the degree of anisotropy only weakly affects source sizes and positions close to the disk center, but has a stronger effect close to the limb. Thus, the radial size (along the $X$-axis) for $\alpha =0.5$ (nearly isotropic scattering) does not decrease towards the limb as fast as in the case with stronger anisotropy $\alpha =0.3$ (Figure \ref{fig:rshift}). This is consistent with the observations of angular broadening of the Crab Nebula \citep[e.g.,][]{1972PASAu...2...86D} by coronal turbulence, which show a preferential elongation along the tangential direction.

Similarly, the interplay between scattering and the focusing effects determine the directivity of the escaping emission. The simulated directivity patterns show that although radio-wave scattering effects lead to large source sizes, the directivity (right panel in Figures \ref{fig:eps08_anis05}-\ref{fig:eps08_anis03}) is predominately into radial direction with half-widths at half-maximum $\simeq 47^o$ and $\simeq 40^o$ for anisotropy $\alpha =0.5$ and $\alpha =0.3$ correspondingly.  These results are different from early results suggesting isotropic directivity due to scattering as reviewed by \cite{1985srph.book..237M}.

\section{Observations of Type III solar radio bursts in the heliosphere: source sizes and decay times}\label{sec:observations}

It is instructive to review observations of solar radio burst source sizes and decay times for comparison with the ray-tracing results. Solar radio bursts are observed over a wide range of frequencies from about $\sim$500~MHz down to $\sim$20~kHz near 1~AU. Therefore, the variation of burst parameters with frequency allows us to diagnose the scattering over a wide range of heliocentric distances.

\begin{figure}[ht!]
    \centering
\includegraphics[width=0.69\textwidth, keepaspectratio=true]{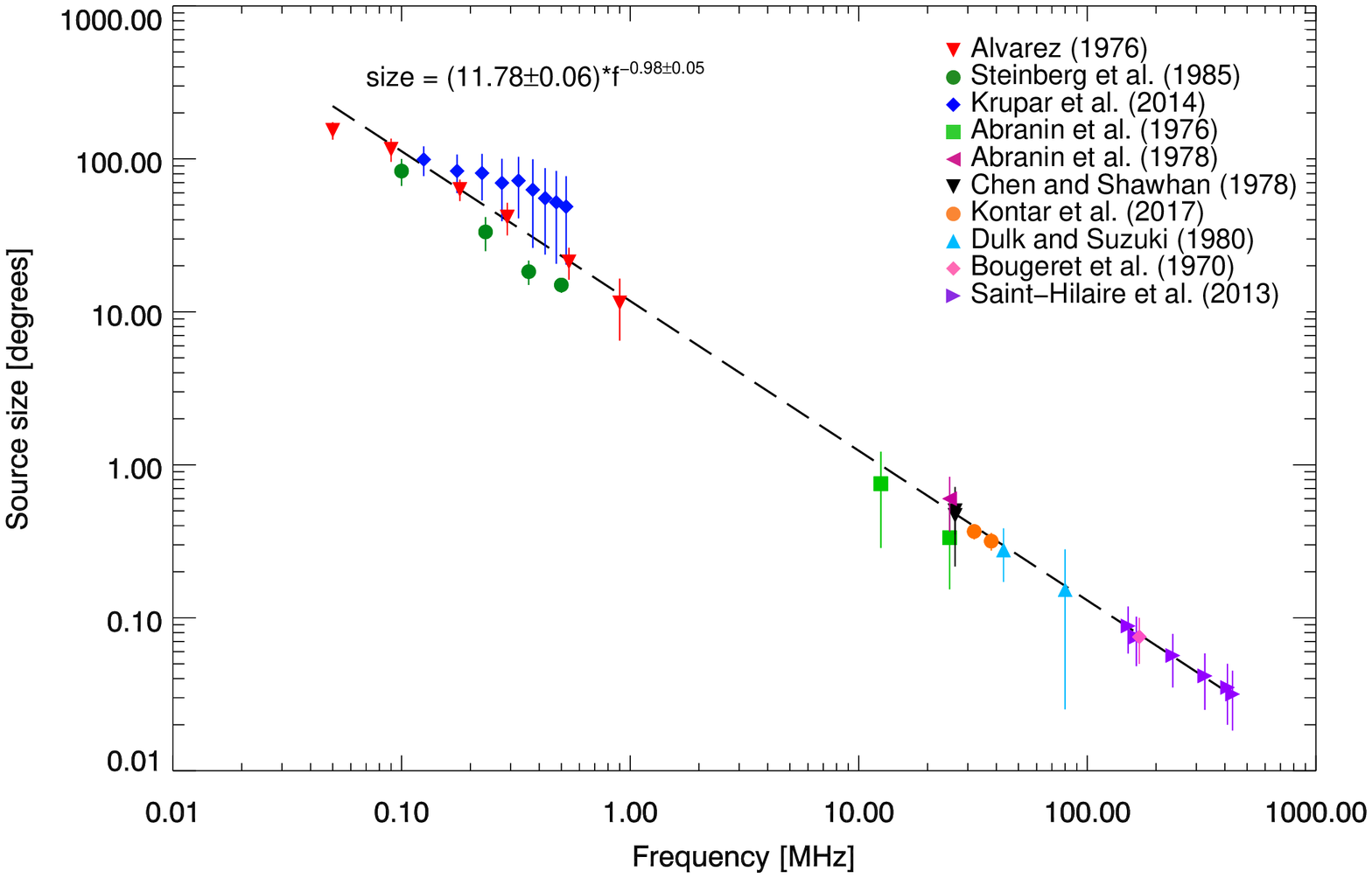}\\
\includegraphics[width=0.69\textwidth, keepaspectratio=true]{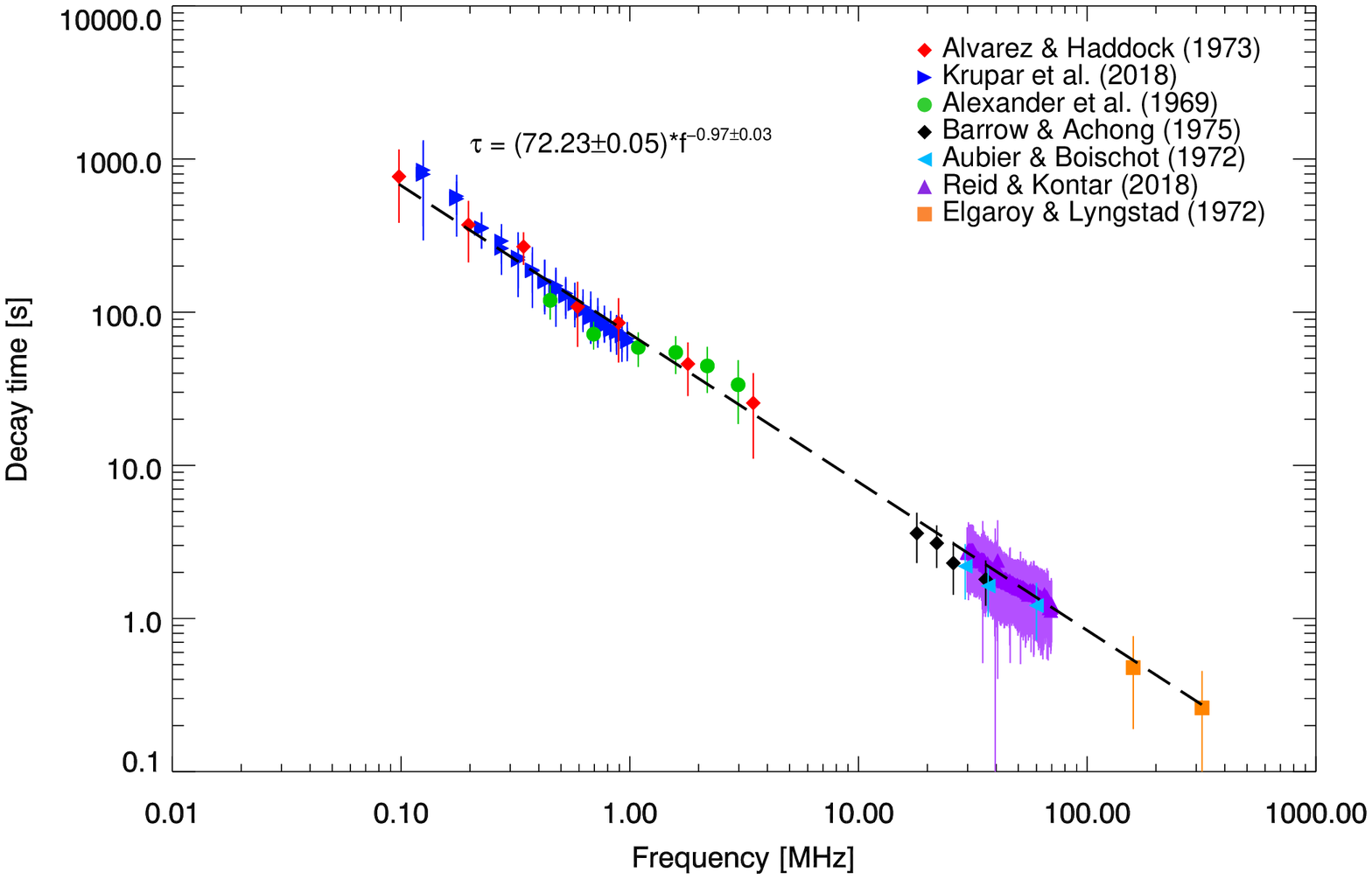}
    \caption{
    \textit{Top:} Source sizes (FWHM; degrees) of type~III solar radio observations versus frequency $f$~(MHz).
     A combination of observations is plotted as indicated by the legend, and a weighted linear fit was applied to the data. The dashed line show the fit given by Equation (\ref{eqn:sizes}).
    \textit{Bottom:} Decay times $\tau$ (defined as the e-folding time in seconds) of Type III solar radio observations, versus frequency $f$ (MHz).  A combination of observations is plotted as indicated by the legend, and a weighted linear fit was applied to the data. The dashed line show the fit given by Equation (\ref{eqn:decays}).
    The standard deviation error bars were calculated from the statistical distribution of the data and measurement errors if reported. }
    \label{fig:type3_data}
\end{figure}

Figure~\ref{fig:type3_data} combines measurements\footnote{The source sizes reported by \cite{1980A&A....88..203D,1985A&A...150..205S} were given as the full width at $1/e$ of the distribution, so the values were recalculated into FWHM values by multiplying by a factor of $\sqrt{\ln 2}$. Measurements above 1~MHz from \cite{2014SoPh..289.4633K} were not plotted as ``the analysis above 1 MHz is perhaps distorted by background signals resulting in increased source sizes'' and thus, the results were deemed unreliable.} by several different authors \citep{1970A&A.....6..406B,1976SvA....19..602A,1976SoPh...46..483A,1978SoPh...57..229A,1978SoPh...57..205C,1980A&A....88..203D,1985A&A...150..205S,2013ApJ...762...60S,2014SoPh..289.4633K,2017NatCo...8.1515K}
over the last fifty years. The Type III source sizes (FWHM; degrees) are for frequencies ranging from $\sim 0.05-500$ MHz. Using a weighted linear fit in log-space, the FWHM depends on the observing frequency ($f$; MHz) as (see Figure~\ref{fig:type3_data})

\begin{equation} \label{eqn:sizes}
\mathrm{FWHM} = (11.8 \pm 0.06) \times f^{-0.98 \pm 0.05} \,\,\, .
\end{equation}

Similarly, a collection of Type III burst decay time measurements
\citep{1969SoPh....8..388A,1972A&A....19..343A,1972A&A....16....1E,1973SoPh...30..175A,1975SoPh...45..459B,2018ApJ...857...82K,2018A&A...614A..69R},
over the frequency range from $\sim 0.1-100$ MHz, is presented in Figure~\ref{fig:type3_data}. The best-fit power-law dependence of the decay time $\tau$ (s) on frequency $f$ (MHz) is

\begin{equation} \label{eqn:decays}
\tau = (72.2 \pm 0.3) \times f^{-0.97 \pm 0.03} \,\,\,\ .
\end{equation}
For comparison, \cite{1950AuSRA...3..541W} derived an expression $\tau=100\times f^{-1}$ for the decay time, based on observations in the frequency range 80~--~120~MHz, while \cite{1973SoPh...30..175A} obtained $\tau=51.29\times f^{-0.95}$ based on observations in the frequency range 50~kHz -- 3.5~MHz, and \cite{1973SoPh...31..501E} obtained $\tau=(2.0\pm1.2)\times100\times f^{-(1.09\pm0.05)}$ based on observations in the frequency range 67~kHz -- 2.8~MHz for $1/e$ decay.

\begin{figure}[pht]
\centering
\includegraphics[width=0.49\linewidth]{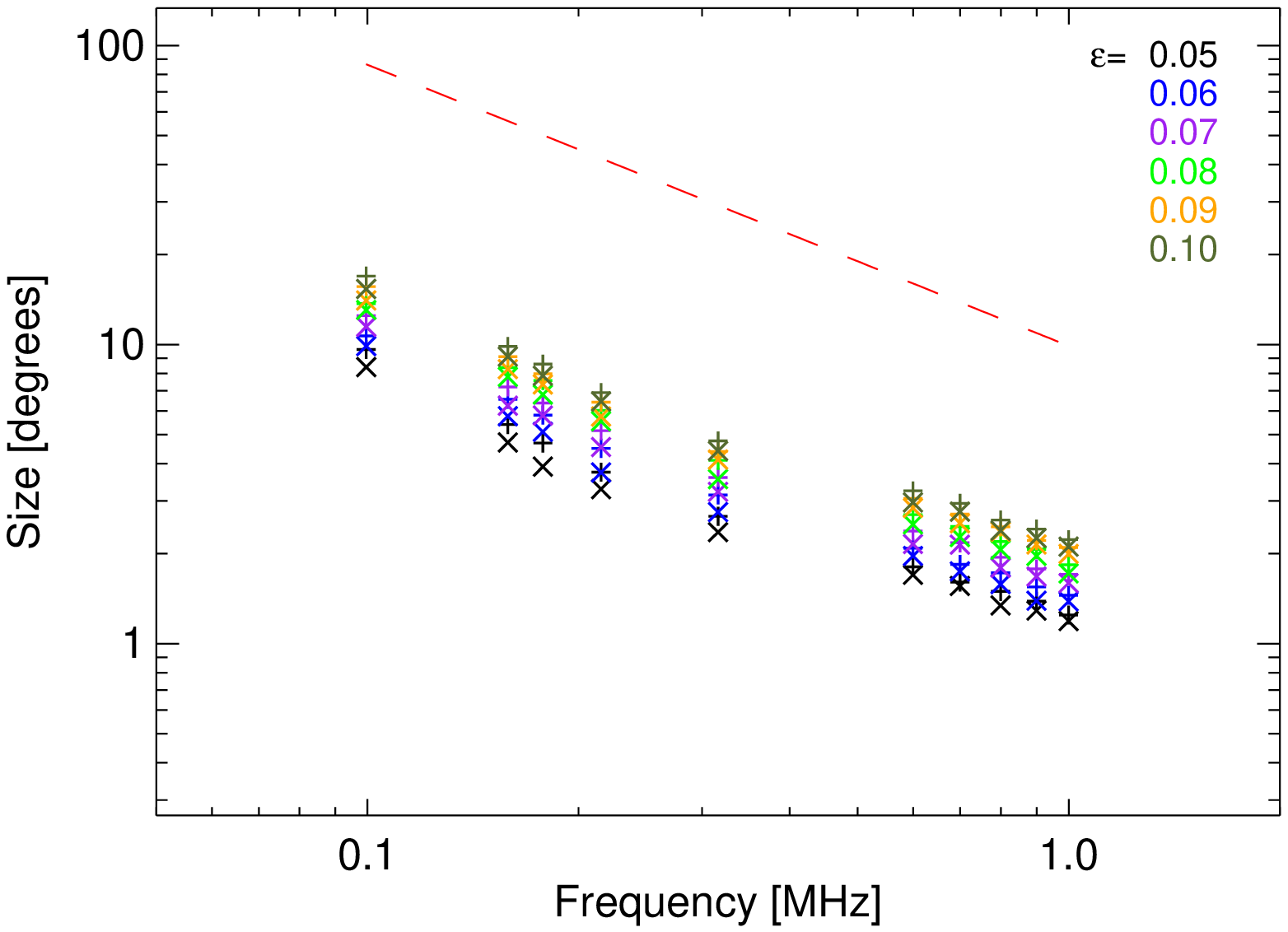}
\includegraphics[width=0.49\linewidth]{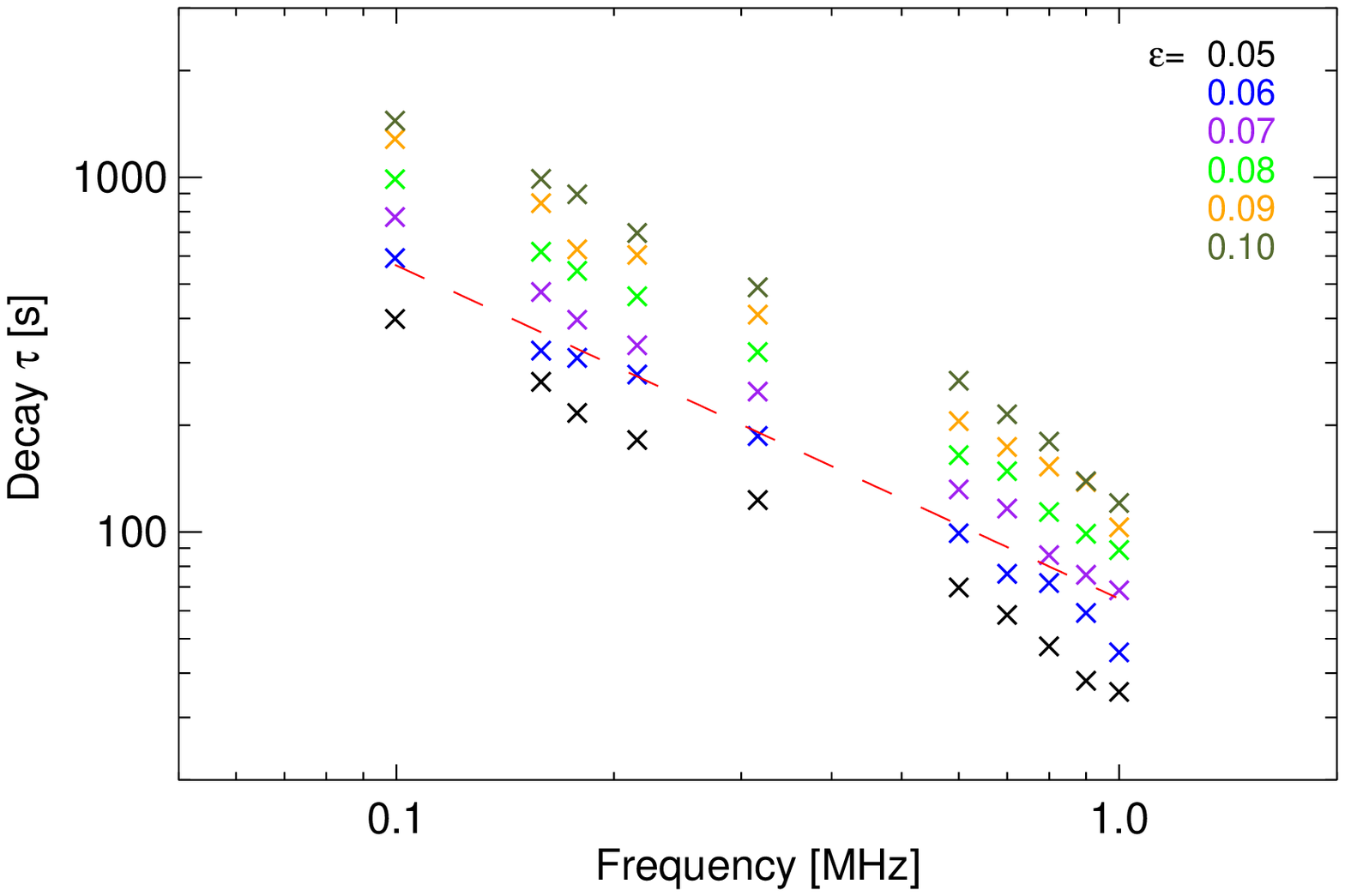}
\caption{\label{fig:size_decay_isotropic} FWHM size (left) and decay time (HWHM) (right) calculated at various frequencies for isotropic scattering and for disk centre source (FWHM$_x$=FWHM$_y$) for frequencies $0.1-1$~MHz. The red dashed line indicates the best fit to the observations from Figure \ref{fig:type3_data}.}
\end{figure}
Figure~\ref{fig:size_decay_isotropic} shows the results of our simulations, assuming isotropic scattering. The decay time agrees within a factor of 2 with that by  \cite{2018ApJ...857...82K}; this difference is likely due to the different numerical schemes used (see the discussion around Equation~(\ref{eq:k2_const})). While a detailed comparison for various anisotropies would require substantial computation effort outside the scope of this work, it is nevertheless clear that isotropic scattering cannot explain the observations. For example, if the level of density fluctuations $\eps$ is chosen to explain the decay times, the predicted source sizes are far too small to explain the observations. Similarly, if the level of isotropic density fluctuations is chosen to match the source sizes, the decay times are too long. Evidently, anisotropic scattering, with a reduced level of scattering along the radial direction, is needed to account for both observed source sizes and decay times.

\section{Summary and Discussion}\label{sec:summary}

Radio emission from solar sources is strongly affected by scattering on small-scale density fluctuations. In general, the observed source sizes and positions, time profiles, and directivity patterns are determined mainly by propagation effects and not by intrinsic properties of the primary source. We have constructed a new model that allows quantitative analysis of radio-wave propagation in a medium that contains an axially symmetric, but anisotropic, scattering component. We have compared the results of numerical simulations using this model with observations of source sizes and time profiles over a wide range of frequencies. Since plasma emission sources with small intrinsic size are observed in type III bursts \citep{2017NatCo...8.1515K,2018SoPh..293..115S}, the observed radio sources are dominated by the scattering, at least at these frequencies. Hence their sizes can be used as diagnostics of radio-wave propagation effects.

In general, a typical source of plasma emission (e.g., Type I, II, III, IV or V solar radio bursts) might have a finite size $FWHM_\text{source}$ defined by the intrinsic size of the region producing the radio emission. The observed FWHM size for such a source is given by $(FWHM_\text{source}^2+FWHM_\text{scat}^2)^{1/2}$, where $FWHM_\text{scat}$ is calculated in this paper. Thus for frequencies around 35~MHz, $FWHM_\text{scat}\simeq 1.1R_{\odot}$, so if the source is substantially smaller than this value, the observed source sizes are dominated by scattering effects. For large sources $\gtrsim 1.1 R_{\odot}$ (i.e., $\gtrsim 18'$), the source sizes due to scattering calculated in this paper can be subtracted in quadrature from the observed source size to give the dimensions of the intrinsic source, corrected for wave propagation effects. However, the size of density fluctuations, and hence the scattering efficiency, can vary appreciably from event to event and from one solar atmosphere region to another, consistent with the considerable variability of the density fluctuation spectrum observed in the solar wind \citep[e.g.,][]{1983A&A...126..293C,1990JGR....9511945M}.

The main result of our work comes from the comparison of the simulation results with combined imaging and time-delay observations. For a given density fluctuation magnitude $\eps$ and outer and inner scales $l_o$, $l_i$, changing the anisotropy parameter $\alpha$ only weakly affects the source size over a broad range of angles near the disk center. (These effects are most noticeable close to the limb, where the anisotropy direction corresponds to the line-of-sight.) However, the time profiles (or, equivalently, the radio pulse expansion along the line of sight) are strongly affected by the value of $\alpha$. Comparison of the simulation results with observations of source size and time delay, both as a function of frequency, suggests that anisotropic density turbulence, with preferential scattering perpendicular to the solar radial direction ($\alpha \simeq 0.3$) is required to account for both the source size and time delay variations at frequencies close to $30$~MHz. In order to explain the Type III observations in the heliosphere between 0.1 and 1~MHz, additional simulations are required. Indeed, the simulations by \citet{2018ApJ...857...82K} demonstrate that although isotropic scattering with $\eps \simeq 0.06-0.07$,  $l_o$ and $l_i$ given by Equation (\ref{eq:q_krupar}) can explain the decay time, the anisotropy of density fluctuations is inadequate to explain the typical source sizes (e.g., Figure \ref{fig:size_decay_isotropic}). The numerical model developed in Section \ref{sec:stochastic-equations} suggests that the anisotropic density fluctuations (lower power in the parallel direction) are required to account for the source sizes and decay times simultaneously. This result requires further computationally-intensive investigations using the method outlined in the paper.

The other interesting result is that the directivity of solar radio bursts is determined by a combination of wave focusing due to large-scale refraction and scattering on small-scale density fluctuations. At the same time, the intrinsic directivity of the source, e.g., the dipole pattern associated with radio emission near the plasma frequency \citep{1970SvA....14..250Z} is quickly lost due to scattering and thus is not evident in observations. Contrary to the results of early simulations \cite[e.g.,][for a review]{1985srph.book..237M}, the resulting directivity appears to have a width of approximately $40$~degrees near 30 MHz. The observed directivity pattern is a combination of the focusing due to large scale refraction and the scattering. The anisotropy of the density fluctuation spectrum plays an important role in governing the emission pattern of solar radio bursts. Therefore, efficient isotropization of radio waves near the emission source does not automatically imply isotropic emission pattern as sometimes assumed.

Free-free absorption appears to have a small or negligible effect for frequencies below $30-50$~MHz. However, the collisions are important for higher frequencies and can determine the time profile. It is also important to note that the stronger the scattering of radio waves, the more pronounced the effect of the free-free absorption. Photons that are strongly scattered are also absorbed stronger and hence produce a weaker contribution to the observed properties.

The effect of radio-wave scattering depends on the radial profiles of the quantities $(\bar{q} \eps ^2 )(r)$ and $\alpha (r)$, representing the size and anisotropy of density fluctuations, respectively. For a decreasing spectrum of electron density fluctuations $S(q)\propto q^{-5/3}$, scattering is most sensitive to the largest $q$ (i.e., the smallest scales) in the inertial range spectrum --- the scale of energy dissipation --- and so provides key diagnostics for the inner scale $l_i(r)$ (Equation~(\ref{eq:q_krupar})). At the same time, conclusions regarding the level of density fluctuations $\eps$ are also dependent on, and so require knowledge of, the outer density scales $l_0$. For example, to explain the observations near 30 MHz, a high level of density fluctuations $\eps =0.8$ is required for the model of $l_o (r)$ adopted, and it is possible that the model $l_o(r)$ is not valid at these frequencies. Comparison between observations and simulations therefore provides a powerful tool with which to infer the radial variation of density fluctuations from the Sun to the Earth, which will be the subject of further work.

\acknowledgements
EPK and NLSJ acknowledges the financial support from the STFC Consolidated Grant ST/P000533/1. XC was supported by the National Natural Science Foundations of China under Grants 11433006 and 11790301. AGE was supported by grant NNX17AI16G from NASA's Heliophysics Supporting Research program. VK acknowledges support by an appointment to the NASA postdoctoral program at the NASA Goddard Space Flight Center administered by Universities Space Research Association under contract with NASA and the Czech Science Foundation grant 17-06818Y. We also acknowledge support from the International Space Science Institute for the LOFAR  \url{http://www.issibern.ch/teams/lofar/} and solar flare \url{http://www.issibern.ch/teams/solflareconnectsolenerg/} teams.

\appendix

\section{Isotropic density fluctuations}\label{app:iso}

For isotropic density fluctuations, due to spherical symmetry,

\[
D_{i j}=D_0\left(\delta _{ij}-\frac{k_i k_j}{k^2}\right) \,\,\, .
\]
Taking the projection of $D _{ij}$ with $\delta _{ij}$ gives

\[
D_{i j}\delta _{ij}=D_0\left(\delta _{ii}-\frac{k_i k_i}{k^2}\right)=2D_0 \,\,\, ,
\]
where the $k_i$ are components of $\vec{k}$ and the summation over repeated indices is implicit. Using the wave number diffusion tensor given by Equation~(\ref{eq:D_iso}),

\begin{equation}\label{eq:d_ij_appendix_iso}
D_{i j}=\frac{\pi \omega_{p e}^{4}}{4 \omega c^{2}} \int q_{i} \, q_{j} \, S(q) \, \delta(\mathbf{q} \cdot \mathbf{k}) \frac{d^{3} q}{(2 \pi)^{3}} \,\,\, ,
\end{equation}
or, in polar coordinates,

\[
D_0=\frac{1}{2}\int q^2 \, S(q) \, \delta(qk\cos\theta) \, 2\pi q^2 \, d\cos\theta \, dq =
\frac{\pi}{k}\int_{0}^{\infty}q^3 \, S(q) \, dq \,\,\, .
\]
Hence one finds that Equation~(\ref{eq:d_ij_appendix_iso}) can be written as

\begin{equation}\label{eq:D_ij_iso_derived}
D_{i j}=\left(\delta _{ij}-\frac{k_i k_j}{k^2}\right) \frac{1}{32\pi}\frac {\omega_{p e}^{4}}{\omega c^{2}k} \int_{0}^{\infty} q^3 \, S(q) \, dq \,\,\, .
\end{equation}
This wave vector diffusion tensor has the same structure as that for Langmuir waves \citep[e.g.,][]{1982PhFl...25.1062G,1991PhFlB...3.1968M,2012ApJ...761..176R}.

\section{Gaussian spectrum of density fluctuations}\label{app:gauss}

Following early works by \cite{1970JGR....75.3715H,1971A&A....10..362S} we assume that the density fluctuations have a Gaussian correlation, so the Gaussian auto-correlation function of the density fluctuations is

\begin{equation}\label{eq:c_r_elipse}
  C(r) = \frac{\langle n(0) \, n(\vec{r})\rangle}{n^2}
=\frac{\langle\delta n^2\rangle}{n^2}
\exp{\left( -\frac{r_\perp^2}{h_\perp^2} -\frac{r_\parallel^2}{h_\parallel^2} \right)} \,\,\, ,
\end{equation}
where $h_\perp^2$ and $h_\parallel$ are the perpendicular and parallel correlation lengths, respectively, and $\langle\delta n^2\rangle$ is the variance of density fluctuations. For isotropic fluctuations,

\begin{equation}\label{eq:gauss_cr}
C(r) = \frac{\langle n(0) \, n(\vec{r})\rangle}{\langle n\rangle^2}
=\frac{\langle\delta n^2\rangle}{n^2}
\exp{\left( -\frac{r^2}{h^2} \right)} \,\,\, ,
\end{equation}
where $h = h_\perp = h_\parallel$ is the correlation length.  The spectrum $S(q)$, defined as

\[
 S(q) =\int C(r) e^{-i\vec{k}\cdot \vec{r}} \, d^3r \,\,\, ,
\]
also has a Gaussian form

\begin{equation}\label{eq:gauss_spectr}
 S(q) = \frac{\langle\delta n^2\rangle }{n^2}\,
 (\pi h^2)^{3/2} \exp \left( - \frac{q^2h^2}{4}\right) \,\,\, ,
\end{equation}
so that the  variance of density fluctuations is

\begin{equation}\label{eq:gauss_norm}
\frac{\langle\delta n^2\rangle}{{n^2}} = \eps ^2 =
\int S(q) \, \frac{d^3q}{(2\pi)^3}=\int_0^{\infty} S(q) \, 4\pi q^2 \, \frac{dq}{(2\pi)^3} \,\,\, .
\end{equation}
Substituting the isotropic Gaussian spectrum~(\ref{eq:gauss_spectr}) into the wave vector diffusion tensor~(\ref{eq:D_ij_iso_derived}), one finds

\[
D_{i j}=\left(\delta _{ij}-\frac{k_i k_j}{k^2}\right) \frac{1}{32\pi } \, \frac {\omega_{p e}^{4}}{\omega c^{2}k} \int_{0}^{\infty}  q^3 \, S(q) \, dq = \left(\delta _{ij}-\frac{k_i k_j}{k^2}\right)\frac{\sqrt{\pi}}{4}\frac{\langle\delta n^2\rangle }{h n ^2}\frac {\omega_{p e}^{4}}{\omega c^{2}k} \,\,\, .
\]
The average wave-number vector $q$, given by Equation~(\ref{eq:q_average}), for the density fluctuation spectrum of Equation~(\ref{eq:gauss_spectr}), is

\begin{equation}\label{eq:q_av_gauss}
\bar{q} =\frac{1} {\eps ^2} \int q \, S(q) \, \frac{d^3q}{(2\pi)^3} = (\pi h^2)^{3/2}  \int_{0}^{\infty} q^3 \,
\exp\left(-\frac{q^2h^2}{4}\right) \, \frac{dq}{(2\pi)^3} = \dfrac{4}{\sqrt{{\pi}}h} \,\,\, ,
\end{equation}
so that the diffusion coefficient $D_{\theta\theta}$ becomes

\begin{equation}\label{eq:D_ij_isogauss}
D_{\theta\theta}=\frac{\pi^2}{4}\frac{\omega_{pe}^4}{\omega c^2k^3} \, \frac{1}{(2\pi)^3}
\int_{0}^{\infty} q^3 \, S(q) \, dq = \frac{\sqrt{\pi}}{4h} \, \frac{\langle\delta n^2\rangle }{n^2}\, \frac{\omega_{pe}^4}{c^2\omega k^3}=\frac{\pi}{16} \, \bar{q}\, \epsilon^2 \frac{\omega_{pe}^4}{c^2\omega k^3} \,\,\, ,
\end{equation}
where $\epsilon^2=\langle\delta n^2\rangle/n^2$.  The angular scattering rate per unit time becomes

\begin{equation}\label{eq:dtheta2_dt_anis}
  \frac{d \langle \theta ^2\rangle}{dt} =2D_{\theta\theta}=\frac{\pi}{8} \, \bar{q}\, \epsilon^2 \, \frac{\omega _{pe}^4}{c^2\omega k^3}
\end{equation}
or, per unit distance, for a photon with group speed $v_{gr}=c^2k/\omega$,

\begin{equation}\label{eq:dTheta2_dx}
\frac{d \langle \theta ^2 \rangle}{dx} = \frac{1}{v_{gr}} \, \frac{d \langle \theta ^2 \rangle}{dt} =\frac{\pi}{8} \, \bar{q}\, \epsilon^2 \, \frac{\omega_{pe}^4}{c^4 k^4}
=\frac{\pi}{8} \, \bar{q}\, \epsilon^2 \frac{\omega _{pe}^4}{(\omega^2- \omega _{pe}^2)^2}
= \frac{\sqrt{\pi}}{2} \, \frac{\epsilon^2}{h} \, \frac{\omega _{pe}^4}{(\omega^2- \omega _{pe}^2)^2} \,\,\, ,
\end{equation}
an expression widely used \citep[e.g.,][]{1952MNRAS.112..475C,1968AJ.....73..972H,1970JGR....75.3715H,1971A&A....10..362S,1997AnGeo..15..387L,2018ApJ...857...82K,2018ApJ...868...79C,2019ApJ...873...48G}
and identical to the expression (27) in \cite{1999A&A...351.1165A} (noting that $h=\sqrt{2}l$ needs to be redefined to obtain $2\eta^*=d \langle \theta ^2\rangle/dt$).

\section{Power-law spectrum of density fluctuations}\label{app:pl}

In-situ observations of density fluctuations suggest an inverse power-law spectrum of density fluctuations $S(q)\propto q^{-(p+2)}$, with the exponent $p$ close to $5/3$ as observed \citep{2013SSRv..178..101A}. This power-law normally holds over a broad inertial range from outer scales $l_0 = 2\pi/q_0$ to inner scales $l_i=2\pi/q_i$ \citep[see, e.g.,][for a review]{2013SSRv..178..101A}:

\begin{equation}\label{eq:pl_spectr}
S(q) = \left\{
                \begin{array}{lll}
                 0, &q>q_0\\
                 \mbox{const} \times q^{-(p+2)}, &q_i<q<q_0 \qquad ,\\
                  0, &q<q_i
                \end{array}
              \right.
\end{equation}
where the constant follows by normalizing the inegrated spectrum to the level of density fluctuations $\langle\delta n^2\rangle$.  Then the spectrum-weighted average wave number $\bar{q}$ (Equation~(\ref{eq:q_average})) becomes \citep{1997AnGeo..15..387L,1999A&A...351.1165A}

\begin{equation}\label{eq:q_pl_spectr}
\bar{q} = \frac{(p-1)}{(p-2)} \, \frac{q_i^{2-p}-q_0^{2-p}}
 {q_i^{1-p}-q_0^{1-p}} \,\,\, .
\end{equation}
This is often simplified further by assuming a large range of wave numbers, so that $q_o \ll q_i$. For example, \cite{2007ApJ...671..894T} and \cite{2018ApJ...857...82K} used Equation~(\ref{eq:q_pl_spectr}) with $p=5/3$ in the limit $q_o \ll q_i$, giving the particularly simple form

\begin{equation}\label{eq:q_pl_spectr-5-3}
\bar{q} \simeq 2 \, q_0^{2/3}q_i^{1/3} =4\pi l_0^{-2/3}l_i^{-1/3} \,\,\, ,
\end{equation}
for which the variance of density fluctuations is

\begin{equation}\label{eq:gauss_norm-app}
\eps ^2 = \int S(q) \, \frac{d^3q}{(2\pi)^3} = \int_0^{\infty} S(q) \, 4\pi q^2 \, \frac{dq}{(2\pi)^3} \,\,\, .
\end{equation}
Then the scattering rate with $\bar{q}$
given by (\ref{eq:q_pl_spectr-5-3}) becomes

\begin{equation}\label{eq:dtheta2_dt_pl}
\frac{d \langle \theta ^2\rangle}{dt} =
2 \frac{\pi}{16} \, \bar{q}\, \epsilon^2 \frac{\omega_{pe}^4}{c^2\omega k^3}=
\frac{\pi^2}{4}  l_0^{-2/3}l_i^{-1/3}
 \epsilon^2 \frac{\omega_{pe}^4}{c^2\omega k^3}
 = \frac{\pi^2}{2}  l_0^{-2/3}l_i^{-1/3}
 \epsilon^2 \frac{\omega_{pe}^4 c}{\omega (\omega ^2 - \omega _{pe}^2)^{3/2}} \,\,\, .
\end{equation}
Expressed as a scattering per unit of length $x$, Equation~(\ref{eq:dtheta2_dt_pl}) is

\begin{equation}\label{eq:dtheta2_dx_pl}
\frac{d \langle \theta ^2\rangle}{dx} = \frac{\pi^2}{2}  l_0^{-2/3}l_i^{-1/3}
 \epsilon^2 \frac{\omega_{pe}^4 }{(\omega ^2 - \omega _{pe}^2)^{2}} \,\,\, .
\end{equation}
This is the expression used by \cite{2008ApJ...676.1338T} and \cite{2018ApJ...857...82K}, but includes an additional factor of $\pi/2$. It also coincides with Equation~(30) from \cite{1999A&A...351.1165A}.

\bibliography{refs}

\end{document}